\documentclass[sigconf]{acmart}
\usepackage{amsmath,amsfonts}
\usepackage{array}
\usepackage[caption=false,font=normalsize,labelfont=sf,textfont=sf]{subfig}
\usepackage{textcomp}
\usepackage{stfloats}
\usepackage{url}
\usepackage{verbatim}
\usepackage{graphicx}
\usepackage{paralist}
\usepackage{amsmath,amsfonts}
\usepackage{array}
\usepackage[caption=false,font=normalsize,labelfont=sf,textfont=sf]{subfig}
\usepackage{textcomp}
\usepackage{stfloats}
\usepackage{url}
\usepackage{verbatim}
\usepackage{graphicx}
\usepackage[ruled,vlined,linesnumbered]{algorithm2e}
\usepackage{algpseudocode}
\usepackage{xcolor}
\usepackage{multirow}
\usepackage{amsfonts}
\usepackage{xspace}
\usepackage{colortbl}
\usepackage{makecell}
\usepackage{microtype}
\usepackage{enumitem}
\usepackage{pifont}
\usepackage{booktabs}
\usepackage[switch]{lineno}

\newcommand{\tool}{\textit{STCLocker}\xspace}
\newcommand{\oracle}{\textit{Deadlock Oracle}\xspace}
\newcommand{\feedback}{\textit{Conflict Feedback}\xspace}
\newcommand{\mutation}{\textit{Conflict-aware Scenario Generation}\xspace}
\newcommand{\dls}{\text{DLS}\xspace}

\AtBeginDocument{%
  }

\setcopyright{none}
\renewcommand\footnotetextcopyrightpermission[1]{}
\copyrightyear{2018}
\acmYear{2018}
\acmDOI{XXXXXXX.XXXXXXX}
\acmConference[Conference acronym 'XX]{Make sure to enter the correct
  conference title from your rights confirmation email}{June 03--05,
  2018}{Woodstock, NY}
\acmISBN{978-1-4503-XXXX-X/2018/06}

\begin{document}

\title{\tool: Deadlock Avoidance Testing for Autonomous Driving Systems}

\author{Mingfei Cheng}
% \email{mfcheng.2022@phdcs.smu.edu.sg}
\affiliation{
  \institution{Singapore Management University}
  \country{Singapore}
}

\author{Renzhi Wang}
% \email{y.zhou@ntu.edu.sg}
\affiliation{
  \institution{University of Alberta}
  \country{Canada}
}

\author{Xiaofei Xie}
% \email{xfxie@smu.edu.sg}
\affiliation{
  \institution{Singapore Management University}
  \country{Singapore}
}

\author{Yuan Zhou}
% \authornote{Corresponding author}
% \email{y.zhou@ntu.edu.sg}
\affiliation{
  \institution{Zhejiang Sci-Tech University}
  \country{China}
}

\author{Lei Ma}
\affiliation{
  \institution{The University of Tokyo}
  \country{Japan}
}
\affiliation{
  \institution{University of Alberta}
  \country{Canada}
}

\renewcommand{\shortauthors}{Mingfei Cheng, Renzhi Wang, Xiaofei Xie, Yuan Zhou and Lei Ma}

\renewcommand{\shortauthors}{Trovato et al.}

\begin{abstract}
Autonomous Driving System (ADS) testing is essential to ensure the safety and reliability of autonomous vehicles (AVs) before deployment. However, existing techniques primarily focus on evaluating 
% safety-critical 
ADS functionalities in single-AV settings. 
As ADSs are increasingly deployed in multi-AV traffic, it becomes crucial to assess their cooperative performance, 
particularly regarding deadlocks, a fundamental coordination failure in which multiple AVs enter a circular waiting state indefinitely, resulting in motion planning failures.
Despite its importance, the cooperative capability of ADSs to prevent deadlocks remains insufficiently underexplored. 
% particularly with respect to deadlocks, a critical cooperative 
% % non-safety-critical 
% property in multi-AV scenarios.
% where multiple AVs mutually wait for each other and are unable to proceed, resulting motion task failures.
% This gap stems from two key difficulties: the lack of effective oracles for detecting deadlock scenarios, and the complexity of constructing scenarios that can trigger deadlocks among multi-AVs.
To address this gap, we propose the first dedicated Spatio-Temporal Conflict-Guided Deadlock Avoidance Testing technique, \tool, for generating DeadLock Scenarios (\dls{}s), where a group of AVs controlled by the ADS under test are in 
a circular wait state. 
% mutually wait for one another in a cyclic dependency. 
\tool consists of three key components: \oracle, \feedback, and \mutation. 
\oracle{} provides a reliable black-box mechanism for detecting deadlock cycles among multiple AVs within a given scenario. 
% without requiring access to the internal logic of the ADS under test. 
\feedback{} and \mutation{} collaborate to actively guide AVs into simultaneous competition over spatial conflict resources (i.e., shared passing regions) and temporal competitive behaviors (i.e., reaching the conflict region at the same time), thereby increasing the effectiveness of generating conflict-prone deadlocks.
% -temporal conflict resources, thereby increasing the likelihood of triggering potential deadlocks.
% Specifically, \feedback{} provides guidance by jointly analyzing spatial conflicts among trajectories and the temporal behaviors of AVs near identified conflict regions. 
% Based on this feedback, \mutation{} modifies both the spatial routes and temporal attributes (i.e., trigger times) of the scenarios to generate more conflict-prone interactions among AVs.
We evaluate \tool on two types of ADSs: \textit{Roach}, an end-to-end ADS, and \textit{OpenCDA}, a module-based ADS supporting cooperative communication. 
Experimental results show that, on average, \tool generates 3.39$\times$ more \dls than the best-performing baseline. By shifting the focus from single-AV evaluation to multi-AV cooperation, our approach offers a novel and realistic perspective for ADS testing, pushing the boundaries of reliability assessment in autonomous driving systems.
\end{abstract}

\maketitle

\section{Introduction}
Autonomous Driving Systems (ADSs) have rapidly advanced in recent years, offering the potential to significantly improve transportation efficiency and urban mobility. As the central control unit of autonomous vehicles (AVs), an ADS is responsible for enabling the vehicle to operate safely without human intervention.
Modern ADSs integrate artificial intelligence algorithms with symbolic reasoning to perceive and understand their environment using data from a variety of sensors, including cameras, LiDAR, radar, and GPS. Based on this perception, the ADS performs key functions such as decision-making, motion planning, and control to ensure safe and efficient driving.
Despite the substantial progress in ADS development over the past decades, ensuring that these systems can consistently meet performance requirements across diverse and unpredictable driving scenarios remains a significant challenge, due to inherent vulnerabilities and limitations~\cite{garcia2020comprehensive}. As a result, comprehensive testing of ADSs across a wide range of conditions is essential before they can be safely deployed in the real world~\cite{lou2022testing}.

% ADS testing can be broadly categorized into \textit{real-world testing} and \textit{simulation-based testing}. While real-world testing is indispensable for validating system performance in actual driving environments, it is often impractical and costly due to the extensive mileage required and the rarity of encountering long-tail events.
% In contrast, simulation-based testing provides a controllable, efficient, and cost-effective approach for both the development and evaluation of ADSs. It enables the systematic creation and assessment of diverse scenarios, including rare or hazardous conditions that are difficult to reproduce in the real world. 

Simulation-based ADS testing has made significant progress in uncovering weaknesses in ADSs. 
Existing efforts primarily fall into two categories: 
(1) reconstructing scenarios based on real-world accidents~\cite{van2015automated,guo2024sovar,tang2024legend}; and 
(2) designing algorithms to generate scenarios that target specific testing objectives, such as inducing collisions~\cite{av_fuzzer,cheng2023behavexplor,icse_samota,haq2023many,huai2023doppelganger,css_drivefuzzer,lu2022learning,tse_adfuzz}, 
violating traffic rules~\cite{sun2022lawbreaker,zhang2023testing,li2024viohawk}, 
or evaluating non-safety-critical properties such as decision optimality~\cite{decictor}.
However, most of existing works only focus on single-vehicle settings, where the ADS is tested within a single autonomous vehicle and other NPC vehicles. 
Such settings fail to capture the complexity of real-world deployments, where multiple AVs operate concurrently and interact dynamically. 
For instance, commercial platforms like Waymo One~\cite{waymo2024sf} and Apollo Go~\cite{apollo_go} have deployed thousands of autonomous vehicles in urban environments. 
These large-scale deployments have occasionally worsened traffic conditions, leading to incidents such as collisions and congestion~\cite{edm2024waymo, bellan2024waymo}, highlighting the need for more comprehensive testing in multi-AV scenarios.

\begin{figure}[!t]
    \centering
    \includegraphics[width=\linewidth]{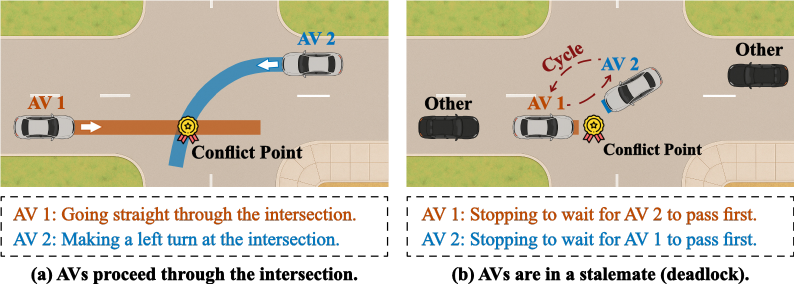}
    \vspace{-20pt}
    \caption{Illustration of a deadlock scenario.}
    \vspace{-15pt}
    \label{fig:intro}
\end{figure}

This paper aims at the testing challenges that arise from the deployment of multiple autonomous vehicles, with a particular focus on \textit{deadlock} scenarios in their cooperative behavior. Effective cooperation among AVs is essential for ensuring safe and efficient interactions in dense traffic environments. 
Deadlocks can occur when two or more AVs, each controlled by the ADS, attempt to pass through a common region but end up waiting for one another to proceed, ultimately resulting in a circular wait.
Fig.~\ref{fig:intro} illustrates an example of a deadlock at an intersection. As shown in Fig.~\ref{fig:intro}(a), both AVs are proceeding through the intersection: AV~1 is going straight, while AV~2 is making a left turn. However, as they approach the shared conflict point, AV~1 detects that AV~2 intends to turn left and decides to stop to let AV~2 pass first. At the same time, AV~2 also stops, waiting for AV~1 to proceed first. This 
circular wait results in a deadlock, as illustrated in Fig.~\ref{fig:intro}(b). Such deadlocks can completely halt traffic flow, block other vehicles and entire road segments, and lead to broader negative impacts on transportation efficiency and societal well-being. Therefore, it is crucial to develop systematic testing techniques to evaluate the deadlock-handling capabilities of ADSs in multi-AV scenarios.

% In real-world deployments, however, multiple AVs often operate concurrently in the same environment, necessitating more comprehensive testing under multi-vehicle scenarios. 
% For example, Waymo One~\cite{waymo2024sf} has deployed over 300 autonomous vehicles as part of its ride-hailing service in San Francisco, highlighting the need to assess ADS behaviors in multi-agent contexts.

% Simulation-based ADS testing has made significant progress, particularly in identifying weaknesses of ADSs. These efforts mainly focus on two directions: (1) reconstructing scenarios from real-world accidents~\cite{van2015automated,guo2024sovar,tang2024legend}; and (2) designing algorithms to generate scenarios that target specific testing objectives, such as inducing collisions~\cite{av_fuzzer,cheng2023behavexplor,icse_samota,haq2023many,huai2023doppelganger,css_drivefuzzer,lu2022learning,tse_adfuzz}, breaking traffic rules~\cite{sun2022lawbreaker,zhang2023testing,li2024viohawk} or evaluating non-safety-critical properties like decision optimality~\cite{decictor}.

% in single-vehicle settings. 
% However, with the growing trend of large-scale ADS deployment, more comprehensive testing is required. For example, Waymo One~\cite{waymo2024sf} has deployed over 300 autonomous vehicles as part of its ride-hailing service in San Francisco. As such, focusing solely on testing ADSs in single-vehicle settings is no longer sufficient to ensure their reliability and safety in real-world, multi-AVs traffic environments.

To the best of our knowledge, 
% there is limited research on evaluating non-safety-critical cooperative abilities of ADSs, specifically, their capability to avoid deadlocks.
there is limited research on evaluating the ADS cooperative abilities among multiple AVs, specifically, the capability to avoid deadlocks~\cite{2011Collision}.
Although some existing ADS testing techniques, such as DoppelTest~\cite{huai2023doppelganger}, started to test scenarios with multiple autonomous vehicles, their objective remains focused on identifying individual weaknesses of the ADS under test in single-AV settings, such as determining its responsibility in violations, rather than evaluating its global cooperation in multi-AV scenarios.

% —often due to safety considerations—
% This cooperative ability is critical for ensuring reliable multi-AVs interactions in dense traffic environments. 
% Specifically, deadlocks can occur when two or more AVs, each controlled by the ADS, attempt to pass through a common region but end up waiting for one another to proceed, ultimately resulting in a circular wait.
% Fig.~\ref{fig:intro} illustrates an example of a deadlock at an intersection. As shown in Fig.~\ref{fig:intro}(a), both AVs are proceeding through the intersection: AV~1 is going straight, while AV~2 is making a left turn. However, as they approach the shared conflict point, AV~1 detects that AV~2 intends to turn left and decides to stop to let AV~2 pass first. At the same time, AV~2 also stops, waiting for AV~1 to proceed first. This 
% % mutual hesitation 
% circular wait results in a deadlock, as illustrated in Fig.~\ref{fig:intro}(b). Such deadlocks can completely halt traffic flow, block other vehicles and entire road segments, and lead to broader negative impacts on transportation efficiency and societal well-being. Therefore, it is crucial to develop systematic testing techniques to evaluate the deadlock-handling capabilities of ADSs in multi-AV scenarios.

To fill this gap, we present the first study on evaluating the cooperative abilities of ADSs in multi-AV settings, with a particular focus on assessing their capability to avoid deadlocks. However, AV deadlock detection poses two major technical challenges.
\ding{182}~First, deadlocks are difficult to identify based solely on temporary stationary behavior without access to high-level semantic intentions. 
ADSs, particularly those based on deep learning models, do not expose internal semantic states, making it challenging to infer the underlying inter-vehicle dependencies necessary for accurate deadlock detection.
\ding{183}~Second, deadlocks are sparsely distributed in the scenario space and typically emerge only under specific interactions, road geometries, and traffic configurations. 
For example, given $N$ vehicles and $M$ possible motion choices per vehicle, the joint configuration space scales exponentially as $N^M$, rendering exhaustive exploration infeasible. The complexity further increases when factoring in road topology and vehicle positions, leading to an even larger and more complex search space.
Therefore, effectively identifying traffic and interaction patterns that are prone to deadlock remains a challenge.

To address these challenges, we propose \tool{}, a novel \textit{Spatio-Temporal Conflict-Guided Deadlock Avoidance Testing} method, designed to effectively identify and generate deadlock scenarios (\dls{}s) in multi-AV environments.
\tool{} consists of three main components: \oracle, \feedback, and \mutation.
To tackle the first challenge~\ding{182}, \oracle provides a reliable mechanism for identifying deadlock cycles among multiple AVs by predicting and abstracting low-level observable behaviors into semantic edges in a wait-for graph. This abstraction facilitates the detection of cyclic dependencies that signify deadlocks. 
To address the second challenge~\ding{183}, we design \feedback{} and \mutation{} to collaboratively guide multiple AVs into deadlock states. 
Specifically, \feedback{} provides guidance by jointly analyzing spatial 
conflict regions 
% conflicts 
among trajectories and the temporal behaviors of AVs near the identified 
% conflict 
regions. 
Guided by this feedback, \mutation{} adjusts both the spatial routes and temporal attributes (i.e., trigger times) of the scenarios to generate more conflict-prone interactions among AVs, thereby increasing the likelihood of inducing deadlocks.
% (1) generate diverse spatial traffic topologies that increase the likelihood of deadlock-prone interactions, and (2) adjust temporal relations among AVs to promote simultaneous competition for shared resources.
% In detail, \feedback evaluates the diversity of traffic topologies and estimates the degree of competition by analyzing AV interactions near shared regions. It returns a score to guide the scenario generation process. \mutation then adaptively adjusts scenario attributes, such as introducing novel traffic topologies to expose new shared resource configurations, or slightly modifying the temporal relations among AVs by adjusting their triggering times. These adaptations increase the likelihood of shared resource contention, thereby amplifying competitive intensity and inducing potential deadlocks.

We evaluate \tool on the high-fidelity simulator CARLA~\cite{dosovitskiy2017carla} with two ADSs: \textit{Roach}~\cite{roach_iccv}, an end-to-end model-based ADS, and \textit{OpenCDA}~\cite{xu2021opencda}, a modular ADS that supports cooperative communication.
We compare \tool against two baselines: \textit{Random}, which randomly generates multi-AV scenarios, and \textit{DoppelTest}~\cite{huai2023doppelganger}, which evaluates ADSs in multi-AV settings with a focus on multiple safety-critical properties.
Experimental results demonstrate that \tool is both effective and efficient in generating \dls{}s, revealing the limited cooperative capabilities of current ADSs in avoiding deadlocks. 
For example, \tool successfully generates a total of 188.2 \dls{}s, while the best-performing baseline detects only 58.8 \dls{}s. Further ablation studies confirm the contribution of each component in \tool.

In summary, this paper makes the following contributions:
\begin{enumerate}[leftmargin=*]
\item We present the first study that systematically evaluates the cooperative capabilities of ADSs in multi-AV settings, with a particular focus on their ability to avoid deadlocks, which is a critical yet underexplored non-safety-critical property.

\item We propose a novel search-based method, \tool, which effectively identifies and generates deadlock scenarios (\dls{}s) through spatial and temporal conflict guidance.

\item We conduct extensive experiments on two distinct types of ADSs to demonstrate the effectiveness of \tool and to reveal significant limitations in their cooperative behavior under potential deadlock conditions. 

% \item Our tool~\cite{ourweb} is released to support future research on cooperative testing of ADSs in multi-AV scenarios.

\end{enumerate}

\section{Background}

\subsection{Autonomous Driving Systems} \label{sec-ads-spec}
Autonomous Driving Systems (ADSs) serve as the decision-making core of autonomous vehicles, generating driving actions based on perceived environmental context. Existing ADSs can generally be categorized into two main paradigms: module-based and end-to-end (E2E). 
Module-based ADSs, such as Baidu Apollo~\cite{apollo}, Autoware~\cite{autoware}, and Pylot~\cite{gog2021pylot}, are composed of distinct components, typically including perception, prediction, planning, and control. Each module is responsible for a specific subtask that collectively ensures safe and effective vehicle operation. 
The perception module processes raw sensor data (e.g., camera images and LiDAR point clouds) to detect and interpret environmental elements such as nearby vehicles, primarily using deep learning (DL) models. 
The prediction module estimates the future motion of surrounding objects based on the perception outputs. 
Given the results of the perception and prediction modules, the planning module generates a safe and feasible trajectory for the vehicle to follow. Finally, the control module translates this trajectory into low-level control commands (e.g., steering, throttle, and braking) to execute the planned motion.
With the rapid advancement of deep learning, end-to-end (E2E) ADSs seek to replace traditional module-based architectures with unified deep learning models. For example, UniAD~\cite{hu2023_uniad} and OpenPilot~\cite{openpilot} employ models that directly map sensor inputs to driving commands, bypassing the need for explicit modular decomposition. Other approaches adopt a hybrid paradigm by replacing only specific components within conventional ADSs. Methods such as PlanT~\cite{Renz2022CORL} and Interfuser~\cite{shao2023safety} map sensor data to planned trajectories, while still relying on conventional rule-based controllers (e.g., PID) for low-level vehicle control. Roach~\cite{roach_iccv}, for instance, replaces the decision-making module, taking outputs from the perception system and generating control commands directly.

\begin{figure*}[!t]
    \centering
    \includegraphics[width=0.8\linewidth]{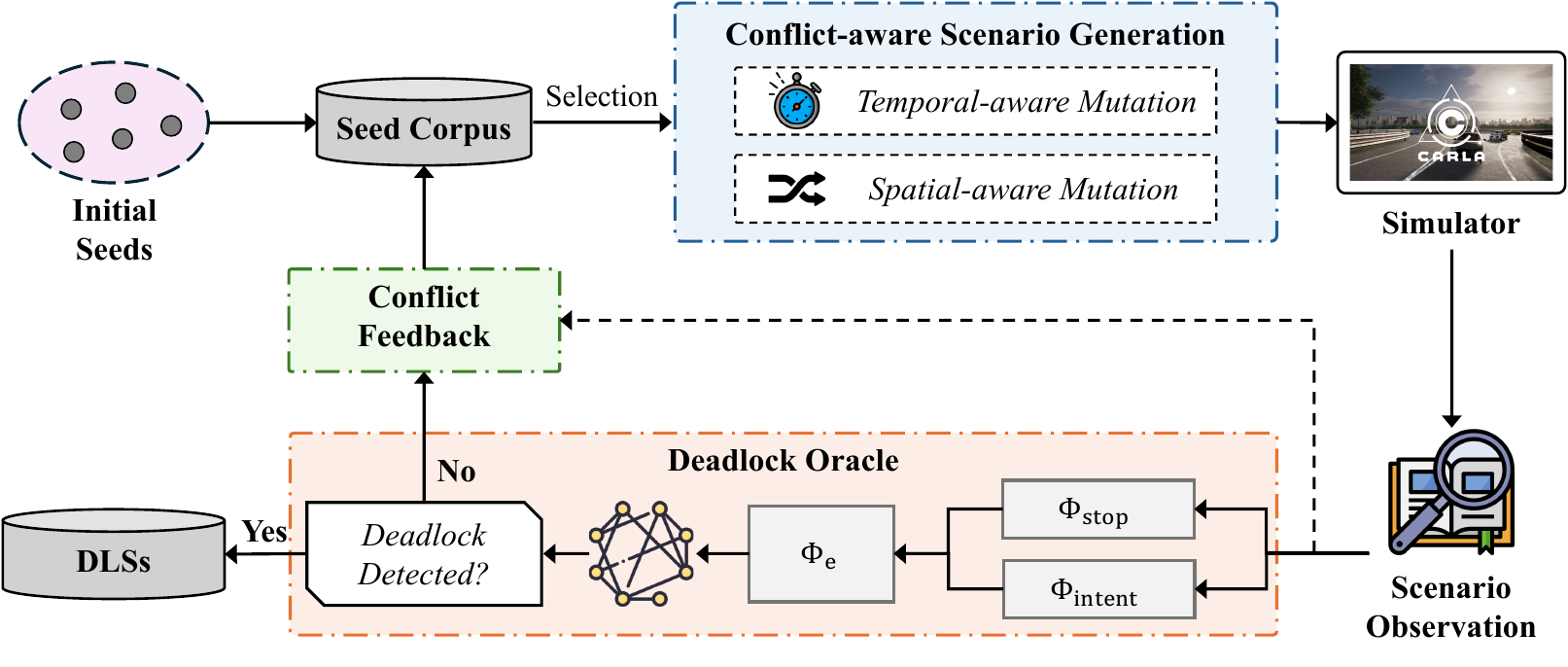}
    \vspace{-15pt}
    \caption{Overview of \tool.}
    \vspace{-10pt}
    \label{fig:overview}
\end{figure*}

\subsection{Scenario} \label{def-scenario}
Testing Autonomous Driving Systems (ADSs) requires a diverse collection of scenarios as testing inputs. Each scenario is typically defined by a set of parameters that specify the environment (e.g., road layout, weather conditions) and traffic participants, including Non-Player Character (NPC) vehicles, pedestrians, and static obstacles. Complex scenarios are often constructed by combining relevant attributes derived from the Operational Design Domains (ODDs)\cite{thorn2018framework}. However, exhaustively covering all attribute combinations is infeasible due to the vastness of the attribute space. Therefore, following prior work\cite{av_fuzzer, cheng2023behavexplor, icse_samota, huai2023doppelganger, lu2022learning, tian2022mosat, css_drivefuzzer}, we select representative subsets of attributes for the purpose of deadlock testing.
Formally, a \emph{scenario} can be represented as a tuple \( S = \{\mathcal{A}, \mathcal{P}\} \), where \( \mathcal{A} = \{ \text{AV}_{1}, \ldots, \text{AV}_{|\mathcal{A}|} \} \) is a finite set of autonomous vehicles (AVs) controlled by the ADS under test, \( \mathcal{P} = \{ P_{1}, \ldots, P_{|\mathcal{P}|} \} \) is a finite set of traffic participants, such as dynamic NPC vehicles.
Specifically, each autonomous vehicle \( \text{AV}_{i} \) is defined as a tuple \( \text{AV}_{i} = \{p_{i}^{\text{start}}, p_{i}^{\text{dest}}, t_{i}^{\text{trigger}} \} \), where \( p_{i}^{\text{start}} \) and \( p_{i}^{\text{dest}} \) denote the starting position and destination, respectively, and \( t_{i}^{\text{trigger}} \) indicates the trigger time at which the vehicle begins to move. Each participant \( P_{i} \in \mathcal{P} \) is represented as a sequence of waypoints \( P_{i} = \{ w_{i}^{1}, \ldots, w_{i}^{N_{i}} \} \), where \( N_{i} \) denotes the number of waypoints. Each waypoint \( w = \{p, \theta, v\} \) encodes the position \( p \), heading angle \( \theta \), and expected speed \( v \).
\emph{Scenario observation} refers to a sequence of scenes that capture the execution of a scenario within a simulation. Formally, given a scenario \( S \), its observation is denoted as \( \mathcal{O}(S) = \{ \mathbf{o}_0, \ldots, \mathbf{o}_T \} \), where \( T \) is the length of the observation and \(\mathbf{o}_t \) represents the scene at timestamp \( t \).
In detail, at timestamp \( t \), the scene observation is denoted as 
\(\mathbf{o}_t = \{ y_{t}^{j} \mid j \in \{\mathcal{A} \cup \mathcal{P} \} \} \), 
where \( y_{t}^{j} \) represents the observation of the \( j \)-th participant. 
Each observation is defined as \( y_{t}^{j} = \{ p_{t}^{j}, \theta_{t}^{j}, v_{t}^{j}, a_{t}^{j} \} \), 
which includes the center position \( p_{t}^{j} \), heading angle \( \theta_{t}^{j} \), velocity \( v_{t}^{j} \), 
and acceleration \( a_{t}^{j} \) at timestamp \( t \). 
By default, we use \( \mathcal{O}(\mathcal{A}) \) to denote the set of AV observations in scenario \( S \), where each \( \mathbf{o}_t \in \mathcal{O}(\mathcal{A}) \) is defined as \( \mathbf{o}_t = \{ y_{t}^{j} \mid j \in \mathcal{A} \} \).

% \(\mathbf{o}_t = \{ y_{t}^{\text{AV}_{1}}, \ldots, y_{t}^{\text{AV}_{|\mathcal{A}|}}, y_{t}^{P_{1}}, \ldots, y_{t}^{P_{|\mathcal{P}|}} \}\),
% where \( y_{t}^{\text{AV}_{i}} \) and \( y_{t}^{P_{j}} \) represent the observations of the autonomous vehicles and the NPC participants, respectively. Each observation \( y_{t}^{j} = \{ p_{t}^{j}, \theta_{t}^{j}, v_{t}^{j}, a_{t}^{j} \} \) denotes the state of a participant \( j \in \{\mathcal{A} \cup \mathcal{P} \} \) at timestamp \( t \), including the center position \( p_{t}^{j} \), heading angle \( \theta_{t}^{j} \), velocity \( v_{t}^{j} \), and acceleration \( a_{t}^{j} \).
% In our setting, we also collect the planning outputs of the ADS at each timestamp \( t \), denoted as \( \tau_{t}^{\text{AV}_{i}} = \{ w_{t}, w_{t+1}, \ldots, w_{t+\Delta t} \} \) for each \( \text{AV}_{i} \), where \( w_{t} \) represents a waypoint and \( \Delta t \) is the planning horizon.

\section{Problem Definition}

This paper aims to evaluate deadlock situations that occur during the deployment of an ADS across multiple autonomous vehicles (AVs) operating in traffic. 
% A deadlock occurs when a set of vehicles mutually block each other's progress, resulting in unexpected traffic congestion. 
Following the deadlock definitions in prior work~\cite{qi2022intersection, pratissoli2023hierarchical}, we define the deadlock scenario (DLS) as follows:

\begin{definition}[Wait-for Graph]
% Given a set of AVs \( \mathcal{A} \) at timestamp $t$, the wait-for graph is denoted as \( G_{t} = (\mathcal{V}, \mathcal{E}_{t}) \). Here, \( \mathcal{V} = \mathcal{A} \) is the set of vertices representing the AVs, and \( \mathcal{E}_{t} \subseteq \mathcal{V} \times \mathcal{V} \) is the set of directed edges, where an edge \( e_{t}^{i \rightarrow j} \in \mathcal{E}_{t} \) indicates that \( \text{AV}_{i} \) is in the wait-for state that stoped and waiting for the move of \( \text{AV}_{j} \) at timestamp $t$.
Given a scenario $S$ and its observation $\mathcal{O}(S)$, the wait-for graph with respect to $\mathbf{o}_t$ is defined as \( G_{t} = (\mathcal{V}, \mathcal{E}_{t}) \). Here, \( \mathcal{V} = \mathcal{A} \) is the set of vertices, each representing an AV, and \( \mathcal{E}_{t} \subseteq \mathcal{V} \times \mathcal{V} \) is the set of directed edges at time \( t \). A directed edge \( e_{t}^{i \rightarrow j} \in \mathcal{E}_{t} \) indicates that \( \text{AV}_{i} \) is waiting for \( \text{AV}_{j} \) to move at timestamp \( t \).
%before it can proceed.
\end{definition}

\begin{definition}[Deadlock Scenario]\label{def:deadlock}
A scenario $S$ is called a deadlock scenario, if there exists a scene $\mathbf{o}_t\in \mathcal{O}(S)$ such that its cooresponding wait-for graph $G_t=(\mathcal{V}, \mathcal{E}_t)$ satisfies \(\{e_t^{i_{1}\rightarrow i_{2}}, e_t^{i_{2}\rightarrow i_{3}} \ldots, e_t^{i_{k}\rightarrow i_{1}}\} \subseteq \mathcal{E}_t\).
% Given a scenario \( S = \{ \mathcal{A}, \mathcal{P} \} \) and its corresponding wait-for graph \( G = (\mathcal{V}, \mathcal{E}) \), there exists a deadlock if \( \exists \mathcal{V}_{i} = \{v_{i_1}, v_{i_2}, \ldots, v_{i_k}\} \subseteq \mathcal{V} \) such that \(\{e_{i_{1}\rightarrow i_{2}}, e_{i_{2}\rightarrow i_{3}} \ldots, e_{i_{k}\rightarrow i_{1}}\} \subseteq \mathcal{E}\).
% Given a scenario \( S = \{ \mathcal{A}, \mathcal{P} \} \) and its corresponding wait-for graph \( G = (\mathcal{V}, \mathcal{E}) \), there exists a deadlock if \( \exists \mathcal{V}_{i} = \{v_{i_1}, v_{i_2}, \ldots, v_{i_k}\} \subseteq \mathcal{V} \) such that \(\{e_{i_{1}\rightarrow i_{2}}, e_{i_{2}\rightarrow i_{3}} \ldots, e_{i_{k}\rightarrow i_{1}}\} \subseteq \mathcal{E}\).
\end{definition}

A deadlock occurs when a group of AVs enters a circular wait, with each vehicle blocked by another in the group. As a result, they occupy shared traffic resources without making progress, potentially causing severe congestion. As illustrated in Fig.~\ref{fig:intro}, such deadlocks can significantly degrade traffic efficiency. The goal of deadlock testing is to generate scenarios that induce deadlock among AVs without violating safety constraints. Notably, deadlocks can occur in both communication-enabled and communication-free settings.

% Once a group of AVs enters a deadlock cycle, they become engaged in a circular wait, where each vehicle is blocked by another in the group. As a result, these AVs occupy traffic resources without making progress, potentially leading to severe congestion. 
% As illustrated in Figure~\ref{}, such deadlocks can significantly degrade traffic efficiency. 
% The primary objective of deadlock testing is to generate scenarios that cause a group of AVs to enter a deadlock state without violating safety constraints. 
% It is important to note that deadlocks can arise both in settings where AVs are capable of communication and in those where no inter-vehicle communication is available.

% The definition poses two main challenges:

% How to model the interaction graph 

% Another challenging thing is how to effectively. 

\begin{algorithm}[!t]
\small
\SetKwInOut{Input}{Input}
\SetKwInOut{Output}{Output}
\SetKwInOut{Para}{Parameters}
\SetKwProg{Fn}{Function}{:}{}
\SetKwFunction{AE}{\textbf{ScenarioGeneration}}
\SetKwFunction{OI}{\textbf{DeadlockOracle}}
\SetKwFunction{FD}{\textbf{ConflictFeedback}}
\SetKwComment{Comment}{\color{blue}// }{}
\Input{
Initial seed corpus $\mathbf{Q}$}
\Output{
Deadlock scenario set $\mathbf{F}$
}
% \Para{
%     Adaptive evolution threshold ${\lambda}_{e}$
% }
$\mathbf{F} \gets \{\}$ \\
\Repeat{given time budget expires}{
\Comment{Select a seed from corpus}
$\{ S, {\phi}_{S} \} \gets \textbf{SeedSelection}(\mathbf{Q})$ \\ 
\Comment{Generate new scenarios}
$S' \gets \AE(S, {\phi}_{S})$ \\
$\mathcal{O}(S') \gets \textbf{Simulator}(S')$ \\
\Comment{Analyze deadlock cycles}
$r \gets \OI(\mathcal{O}(S'))$  \\
% determine mccs
\eIf{$r$ is \textit{Fail}}{
    $\mathbf{F} \gets \mathbf{F} \cup \{S'\}$ \Comment{Update failure sets}
    % $\textbf{continue}$
}{
    \Comment{Update seed corpus}
    ${\phi}_{S'} \gets \FD(\mathcal{O}(S'))$ \\
    \If{${\phi}_{S'} < {\phi}_{S}$}{
        $\mathbf{Q} \gets \mathbf{Q} \cup \{S'\}$ \Comment{Update seed corpus}
    }
}
}

\Return $\mathbf{F}$
\caption{Workflow of \tool}
\label{algo:workflow}
\end{algorithm}

\section{Methodology}
\subsection{Overview}
% \cmf{Add figures about topology comparison.}
Fig.~\ref{fig:overview} presents a high-level overview of \tool{}, which generates DeadLock Scenarios (\dls{}s) from initial seed scenarios. 
\tool{} consists of three core components: \oracle{}, \feedback{}, and \mutation{}. 
Specifically, \oracle{} constructs a wait-for graph based on observed AV behaviors and analyzes it to detect the presence of deadlock cycles. 
\feedback{} guides the testing process by analyzing spatial trajectory patterns and temporal interactions among the group of AVs. 
Based on the computed feedback scores, \mutation{} adaptively selects and applies either spatial-aware mutation or temporal adjustment strategies to increase the likelihood of inducing deadlocks.

Algorithm~\ref{algo:workflow} outlines the core workflow of \tool{}, which follows a classical search-based testing paradigm. 
It maintains a seed corpus that is progressively enriched with high-potential scenarios to facilitate the efficient discovery of \dls{}s. 
Specifically, given an initial seed corpus \( \mathbf{Q} \), \tool{} begins by initializing an empty set \( \mathbf{F} \) to store the detected \dls{}s (Line~1). 
The testing process then proceeds iteratively until the specified time budget is exhausted, after which the set \( \mathbf{F} \) is returned (Line~13). 
In each iteration, \tool{} first selects a seed \( S \), along with its associated feedback score \( \phi_S \), from the seed corpus \( \mathbf{Q} \) (Line~3). 
Then, \mutation{} generates a new scenario \( S' \) by modifying spatial and temporal attributes of \( S \) based on \( \phi_S \), with the goal of increasing inter-vehicle competitiveness and the likelihood of inducing deadlocks (Line 4).
The scenario \( S' \) is then executed in the simulator to obtain its observation \( \mathcal{O}(S') \) (Line~5). 
Based on this observation, \oracle{} constructs a wait-for graph and determines whether \( S' \) results in a failure, i.e., whether it contains a deadlock cycle, and returns the outcome \( r \) (Line~6). 
If \( r \) is marked as \textit{Fail}, the scenario \( S' \) is added to the \dls{} set \( \mathbf{F} \) (Lines~7-8). 
If no failure is detected, \feedback{} evaluates the scenario \( S' \) by computing a feedback score \( \phi_{S'} \), which reflects the degree of spatial conflicts and temporal interaction among the AV group (Lines~9-10). 
If \( \phi_{S'} < \phi_S \), the new seed \( S' \) is retained in the corpus \( \mathbf{Q} \) (Lines~11-12), indicating that \( S' \) has a higher potential to evolve into a deadlock scenario compared to the original seed. 
The process repeats until the time limit is reached, after which the final set of detected \dls{}s \( \mathbf{F} \) is returned (Line~14).

\subsection{\oracle}

As defined in Definition~\ref{def:deadlock}, a deadlock is identified by detecting a scene $\mathbf{o}_t$ whose wait-for graph $G_t=(\mathcal{V}, \mathcal{E}_t)$ exists a cycle \((AV_{i_1}, e_t^{i_{1}\rightarrow i_{2}}, AV_{i_2}, e_t^{i_{2}\rightarrow i_{3}}, AV_{i_3}, \ldots, AV_{i_k}, e_t^{i_{k}\rightarrow i_{1}}, AV_{i_1})\).

% a cycle in the wait-for graph \( G = (\mathcal{V}, \mathcal{E}) \), where each vertex \( v_{i} \in \mathcal{V} \) represents an autonomous vehicle \( \text{AV}_{i} \in \mathcal{A} \), and each directed edge \( e_{i \rightarrow j} \in \mathcal{E} \) indicates that \( \text{AV}_{i} \) is waiting for \( \text{AV}_{j} \) to proceed before it can continue its own motion.

Therefore, given a scenario observation \( \mathcal{O}(S) \),  we need to extract the corresponding Wait-for Graph for each time, with a particular focus on identifying the wait-for edges associated with each AV.
However, given a scenario observation, we can only access low-level outputs of the ADS, such as trajectories and vehicle states, which makes it challenging to infer high-level wait-for intentions. 
To address this, given an autonomous vehicle \( \text{AV}_{i} \), we estimate its wait-for edge \( e_{i \rightarrow j} \) with respect to another vehicle \( \text{AV}_{j} \) by evaluating whether the following condition holds:
\[
\Phi_{\text{edge}}(e_{i \rightarrow j}) = 
\Phi_{\text{stop}}(\text{AV}_{i}) \wedge \Phi_{\text{intent}}(\text{AV}_{i}, \text{AV}_{j})
\]
Here, \( \Phi_{\text{stop}} \) determines whether \( \text{AV}_{i} \) is in a stationary state, and \( \Phi_{\text{intent}} \) predicts whether \( \text{AV}_{j} \) is likely to move into the intended path of \( \text{AV}_{i} \), thereby preventing it from proceeding.

The first condition \( \Phi_{\text{stop}} \) can be easily determined by monitoring the speed of the vehicle by:
\[
\Phi_{\text{stop}}(\text{AV}_{i}) = 
\begin{cases}
\texttt{True}, & \text{if } v_{i}(t') < \epsilon, \, \forall t' \in [t_i^{stop} + \Delta t, t_i^{stop}] \\
\texttt{False}, & \text{otherwise}
\end{cases}
\]
where \( v_{i}(t') \) denotes the speed of \( \text{AV}_{i} \) at time \( t' \), \( \epsilon \) is a small velocity threshold (e.g., 0.01 m/s), and \( \Delta t \) is the minimum duration for which the AV must remain below this threshold to be considered stationary.

To evaluate the second condition \( \Phi_{\text{intent}} \), we examine whether there is an intention conflict between \( \text{AV}_{i} \) and \( \text{AV}_{j} \) during the time interval \( [ t_i^{\text{stop}} - \Delta t,\; t_i^{\text{stop}} ] \), where \( t_i^{\text{stop}} \) denotes the time at which \( \text{AV}_{i} \) comes to a complete stop. Specifically, given historical motion observations, we first predict the intended trajectory of \( \text{AV}_{i} \) over the pre-stop interval, denoted by \( \tau_i^{\text{pre}} \), and the future trajectory of \( \text{AV}_{j} \), denoted by \( \tau_j^{\text{pred}} \), using a Kalman Filter-based prediction approach~\cite{welch1995introduction, kumar2017traffic, emami2019using}.
Then, the condition \( \Phi_{\text{intent}} \) is evaluated as follows:
\[
\Phi_{\text{intent}}(\text{AV}_{i}, \text{AV}_{j}) = 
\begin{cases}
\texttt{True}, & \text{if } \text{Conflict}(\tau_i^{\text{pre}}, \tau_j^{\text{pred}}) = \texttt{True} \\
\texttt{False}, & \text{otherwise}
\end{cases}
\]
Here, \( \text{Conflict}(\tau_i^{\text{pre}}, \tau_j^{\text{pred}}) \) returns \texttt{True} if the two sets of trajectories overlap in both space and time, suggesting a potential interaction that may have contributed to \( \text{AV}_{i} \)'s decision to stop.

Finally, by iterating over all autonomous vehicles in \( \mathcal{A} \) and evaluating the condition \( \Phi_{\text{edge}}(e_{i \rightarrow j}) \) for each vehicle pair, we construct the complete wait-for graph \( G \). A deadlock can then be detected by checking whether \( G \) contains at least one cycle, i.e., \( \Phi_{\text{cycle}}(G) = \texttt{True} \).

\subsection{\feedback}
The feedback component aims to effectively guide the generation of scenarios that are likely to induce mutual exclusion conditions. For example, if two ego vehicles have intersecting conflict regions in their planned trajectories, a deadlock is more likely to occur. In addition to spatial factors, temporal attributes are also critical. The vehicles should approach the conflict region simultaneously or within a short time window to increase the likelihood of triggering a deadlock.
Therefore, we design the feedback by considering  two complementary aspects: 
(1) the \textit{Spatial-Conflict Score}, which captures the spatial conflict relationships among AVs' trajectories; and 
(2) the \textit{Temporal-Conflict Score}, which reflects the temporal competitiveness between AVs that are near conflict regions.

\begin{figure}[!t]
    \centering
    \includegraphics[width=\linewidth]{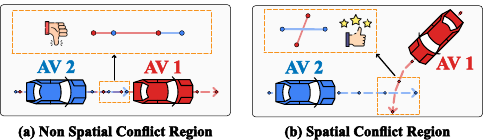}
    \vspace{-15pt}
    \caption{Illustration of spatial conflict regions.}
    \vspace{-15pt}
    \label{fig:spatial_region}
\end{figure}

\subsubsection{Spatial-Conflict Score}\label{sec:st_score} 
% The \textit{spatial topology score} identifies novel traffic topology patterns that are underrepresented in the seed corpus, thereby increasing the diversity of explored scenarios and the chance of discovering deadlocks. Specifically, 
Given a scenario observation \( \mathcal{O}(S) \), we compute the spatial conflict point set \( \mathcal{C}_S \), defined as the set of pairwise intersections between trajectory segments, excluding fully overlapping cases. 
Specifically, the observed trajectory of AV$_i$
% the \( i \)-th AV 
is represented as a sequence of line segments \( \tau_{S}^{i} = \{ \ell_{i,k} \mid k = 1, \ldots, N_{i} - 1 \} \), 
where \( N_i \) is the number of trajectory points, and each segment \( \ell_{i,k} = (\mathbf{p}_{i,k}, \mathbf{p}_{i,k+1}) \) connects two consecutive points. 
Here, \( \mathbf{p}_{i,k} \in \mathbb{R}^2 \) denotes the \( k \)-th spatial point on the trajectory of AV$_i$.
% \( i \). 
We consider two segments \( \ell_{i,k} \) and \( \ell_{j,m} \) to be in spatial conflict if they intersect geometrically and are not aligned in the same direction. 
As illustrated in Fig.~\ref{fig:spatial_region}(a), when AV~1 follows AV~2 along an identical path segment with the same direction, the leading vehicle is unlikely to wait for the following vehicle, as they are not competing for the same conflict region. 
In contrast, Fig.~\ref{fig:spatial_region}(b) demonstrates a case where the two AVs approach the same region from different directions, increasing the likelihood of wait-for behavior due to mutual spatial contention. Formally, a conflict is recorded if:
\[
\ell_{i,k} \cap \ell_{j,m} \neq \emptyset 
\quad \text{and} \quad 
ID(\ell_{i,k})\neq  ID(\ell_{j,m}),
% \angle(\ell_{i,k}, \ell_{j,m}) > \theta,
\]
where $ID(\ell_{i,k})$ denotes the ID the of the lane that $\ell_{i,k})$ belongs to.
% where \( \angle(\ell_{i,k}, \ell_{j,m}) \) denotes the angle between the direction vectors of the two segments, and \( \theta \) is a small threshold (e.g., \( 5^\circ \)) used to exclude near-parallel or collinear cases. 
This ensures that non-competitive interactions such as following or coasting in the same direction are not counted as spatial conflicts. 
The total number of such conflict points, denoted as \( |\mathcal{C}_S| \), provides a measure of the spatial conflict degree in scenario \( S \). 
This value is normalized with respect to the total number of trajectory segments across all AVs and is used as the spatial component of the feedback score:

\[
\phi_S^{\text{spatial}} = 1 - \frac{|\mathcal{C}_S|}{\sum_{i=1}^{|\mathcal{A}|} (N_i - 1)}.
\]
A lower value of \( \phi_S^{\text{spatial}} \in [0, 1] \) indicates that scenario \( S \) exhibits stronger spatial competition, with more frequent trajectory intersections among AVs, suggesting a higher likelihood of potential deadlocks.

\subsubsection{Temporal-Conflict Score}  

For each conflict region \( R \in \mathcal{C}_S \), where \( \mathcal{C}_S \) is the conflict point set obtained in the previous step, we directly estimate the arrival time \( t_R^{\text{AV}_i} \) and speed \( v_R^{\text{AV}_i} \) for each vehicle \( \text{AV}_i \) that passes through region \( R \), based on the scenario observation \( \mathcal{O}(S) \). 
These values can be obtained by identifying the scene timestamp at which the vehicle is closest to the conflict region.
Therefore, for each conflict region \( R \in \mathcal{C}_S \), we compute a temporal conflict score \( \phi_{R} \) as:

\[
\phi_{R} = \min_{\substack{i, j \in \mathcal{A}_{R} \\ i \neq j}} \left( \left| t_R^{i} - t_R^{j} \right| + v_R^{i} + v_R^{j} \right),
\]
where \( \mathcal{A}_{R} \) denotes the set of AVs that pass through the conflict region \( R \), 
\( t_R^{i} \) is the estimated arrival time of \( \text{AV}_i \), and \( v_R^{i} \) is its corresponding speed at region \( R \). 
This score captures the temporal proximity and velocity of competing AVs at the same region, with lower values indicating stronger temporal competition. Then, the final temporal-conflict score for scenario \( S \) is computed as the minimum conflict score across all conflict regions, normalized into the range \( [0, 1] \):

\[
\phi_S^{\text{temporal}} = \text{clip} \left( \frac{\min_{R \in \mathcal{C}_S} \phi_{R}}{N_{\text{ti}}},\ 0,\ 1 \right),
\]
where \( N_{\text{temporal}} \) is a scaling factor used to normalize the score. 
A lower value of \( \phi_S^{\text{temporal}} \) indicates tighter temporal overlap and stronger competition among AVs at spatial conflict regions, particularly at lower speeds, suggesting a higher potential for wait-for behaviors and deadlocks.

% The temporal competition score for \( \text{AV}_i \) is defined as the minimum temporal distance to any other \( \text{AV}_j \) at a shared conflict region \( R \in \mathbf{R}_{ij} \), weighted by its own speed:

% \[
% \phi_{\text{AV}_i} = \min_{j \neq i} \min_{R \in \mathbf{R}_{ij}} \left( \left| t_R^{\text{AV}_i} - t_R^{\text{AV}_j} \right| + v_R^{\text{AV}_i} \right),
% \]

% where \( \mathbf{R}_{ij} \subseteq \mathcal{C}_S \) denotes the set of conflict regions shared between \( \text{AV}_i \) and \( \text{AV}_j \). 
% The temporal interaction score for the scenario \( S \) is then defined as the most competitive (i.e., minimum) score among all AVs, normalized into the range \([0, 1]\):

% \[
% \phi_S^{\text{ti}} = \text{clip} \left( \frac{\min_{i \in \mathcal{A}} \phi_{\text{AV}_i}}{N_{\text{ti}}},\ 0,\ 1 \right),
% \]

% where \( N_{\text{ti}} \) is a scaling factor that normalizes the score. 
% A lower value of \( \phi_S^{\text{ti}} \) indicates tighter timing overlap and stronger temporal competition among AVs at shared regions, particularly at reduced speeds, which suggests a higher likelihood of wait-for behaviors and deadlocks.

\subsubsection{Overall Feedback Score}\label{sec:overall_feedback_score}
Finally, the overall feedback score \( \phi_{S} \in [0, 1] \) for a given scenario \( S \) is calculated as:
\[
\phi_{S} =  \alpha \phi^{\text{spatial}}_{S} + (1-\alpha) \phi^{\text{temporal}}_{S}
\]
which combines the spatial-conflict score \( \phi^{\text{spatial}}_{S} \) and the temporal conflict score \( \phi^{\text{temporal}}_{S} \). The weighting factor \( \alpha \in [0, 1] \) balances the contribution of each component. \tool{} aims to minimize \( \phi_{S} \), thereby guiding the scenario generation process toward cases that are more likely to induce deadlocks.

\begin{algorithm}[!t]
\small
\SetKwInOut{Input}{Input}
\SetKwInOut{Output}{Output}
\SetKwInOut{Para}{Parameters}
\SetKwProg{Fn}{Function}{:}{}
% \SetKwFunction{MM}{\textbf{Mutation}}
\SetKwFunction{MA}{\textbf{TemporalMutation}}
\SetKwFunction{MO}{\textbf{SpatialMutation}}
\SetKwFunction{Mu}{\textbf{ScenarioGeneration}}
\SetKwComment{Comment}{\color{blue}// }{}
\Input{
Scenario $\mathcal{S} = \{ \mathcal{A}, \mathcal{P}\}$\\
Feedback score $\phi_{S}$ \\
Conflict points set $\mathcal{C}_{S}$
}
\Output{
Mutated scenario $S' = \{ \mathcal{A}', \mathcal{P}' \}$
}
\Para{
    Maximum AV Capability $N_{\mathcal{A}}$ \\
    % Maximum NPC Participants Capability $N_{\mathcal{P}}$ \\
    Local search bugdet $N_{local}$
}

\Fn{\Mu{}}{
    % \eIf{random() $> \phi_{S}$}{
    $\mathcal{A}' \gets \MA(\mathcal{A}, \mathcal{C}_{S})$ \\
    \If{$\mathcal{A}'=\emptyset$ $\vee$ \text{random()} $< \phi_{S}$}{
        $\mathcal{A}', \mathcal{P}' \gets \MO(\mathcal{A})$ \\
    }
    $S' \gets \{ \mathcal{A}', \mathcal{P}' \}$ \\
    \Return $S'$
}

\Fn{\MA{$\mathcal{A}$, $\mathcal{C}_{S}$}}{
    \If{$|\mathcal{C}_{S}| = \emptyset$}{
        \Return $\emptyset$
    }
    
    $\mathcal{A}' \gets \mathcal{A}$ \Comment{Copy $\mathcal{A}$ to $\mathcal{A}'$}
    
    $R \gets \text{RandomSelect}(\mathcal{C}_{S})$ \\
    \For{$(\text{AV}_{i}, \text{AV}_{j}) \in \text{RandomPair}(\mathcal{A}_{R})$}{
        $\Delta t'_{R} \gets t_{R}^{\text{AV}_{i}} - t_{R}^{\text{AV}_{j}}$ \\
        \Comment{Update trigger time in $\mathcal{A}'$}
        ${t'}_{i}^{trigger} = \text{max}\{0, t_{i}^{trigger} - \frac{\Delta t'_{R}}{2} \}$ \\
        ${t'}_{j}^{trigger} = \text{max}\{0, t_{j}^{trigger} + \frac{\Delta t'_{R}}{2} \}$ \\
    }
    \Return $\mathcal{A}'$
    % \For{$(\text{AV}_{i}, \text{AV}_{j}) \in \text{RandomPair}(\mathcal{A}')$}{
    %     \Comment{Calculate from observation $\mathcal{O}(S)$}
    %     $\mathbf{R}_{ij} \gets \text{Line}(\mathbf{Y}^{\text{AV}_i}) \cap \text{Line}(\mathbf{Y}^{\text{AV}_j})$ \\
    %     \If{$\mathbf{R}_{ij} \neq \emptyset$}{
    %         $R \gets \text{RandomSelect}(\mathbf{R}_{ij})$ \\
    %         $\Delta t'_{R} \gets t_{R}^{\text{AV}_{i}} - t_{R}^{\text{AV}_{j}}$ \\
    %         \Comment{Update trigger time in $\mathcal{A}'$}
    %         ${t'}_{i}^{trigger} = \text{max}\{0, t_{i}^{trigger} - \frac{\Delta t'_{R}}{2} \}$ \\
    %         ${t'}_{j}^{trigger} = \text{max}\{0, t_{j}^{trigger} + \frac{\Delta t'_{R}}{2} \}$ \\
    %         \Return $\mathcal{A}'$
    %     }
    % }
    % \Return $\mathcal{A}'$
}

\Fn{\MO{$\mathcal{A}$}}{
    $\mathcal{A}' \gets \mathcal{A}$ \Comment{Copy $\mathcal{A}$ to $\mathcal{A}'$} 
    
    \If{$|\mathcal{A}'| \geq N_{\mathcal{A}} $}{
        % \Comment{Simplify scenario}
        $\mathcal{A}' \gets \text{RandomRemove}(\mathcal{A}', \text{min}=0, \text{max}=N_{\mathcal{A}}-2)$
    }
    {
        \Comment{Local search best config}
        $\text{AV}_{best} \gets \text{RandomCreate()}$ \\
        $\phi_{best}^{\text{spatial}} \gets \text{SpatialScore}(\mathbf{Q}, \mathcal{A'} \cup \{ \text{AV}_{best} \})$ \\
        \For{$i < N_{local}$}{
            $\text{AV}_{tmp} \gets \text{RandomCreate()}$ \\ 
             $\phi_{tmp}^{\text{spatial}} \gets \text{SpatialScore}(\mathcal{A'} \cup \{\text{AV}_{tmp} \})$ \\
            \If{$ \phi_{tmp}^{\text{spatial}} < \phi_{best}^{\text{spatial}}$}{
                $\phi_{best}^{\text{spatial}} \gets \phi_{tmp}^{\text{spatial}}$ \\
                $ \text{AV}_{best} \gets \text{AV}_{tmp}$
            }
        
    }
        $\mathcal{A}' \gets \mathcal{A}' \cup \{ \text{AV}_{best} \}$
        
    }

     \Comment{Mutate NPC participants } 
     $\mathcal{P}' \gets  \text{WaypointMutator}(\mathcal{P})$
    
    \Return $\mathcal{A}'$, $\mathcal{P}'$
}
\caption{Algorithm of Scenario Generation}
\label{algo:mutation}
\end{algorithm}

\subsection{\mutation}

To improve the testing performance of \tool{}, 
we need to design scenarios where the AVs have shared regions and arrive at the shared regions simultaneously. 
Therefore,  
we design \mutation{} with two key components: \textit{Temporal-aware Mutation} and \textit{Spatial-aware Mutation}. These components aim to effectively improve feedback scores (both spatial and temporal) by guiding mutations toward promising directions, rather than relying on arbitrary mutations that often yield low scores.

The \textit{Temporal-aware Mutation} adjusts fine-grained attributes of AVs, mainly their start times or initial delays, aiming to encourage the AVs to arrive the shared regions simultaneously.
% to encourage tighter timing overlaps, thereby increasing the likelihood of temporal competition at shared regions.
The \textit{Spatial-aware Mutation}, on the other hand, aims to generate new scenarios with 
% increased potential for spatial
more 
conflict regions among AVs by altering their initial positions, routes, or spawn locations to promote topological convergence. 

Algorithm~\ref{algo:mutation} outlines the scenario generation process, which takes as input the parent scenario \( S \), its feedback score \( \phi_{S} \), and the corresponding conflict point set \( \mathcal{C}_{S} \), and returns a newly generated scenario \( S' \) (Lines~5-6). 
Overall, the generation process first attempts a fine-grained \MA (Lines 7-16) to adjust timing attributes of AVs (Line~2). If the temporal mutation returns an empty set, indicating that the parent scenario contains no conflict regions as determined in Lines~9-10, the process falls back to a spatial mutation (Lines~3-4). Additionally, if the feedback score \( \phi_{S} \) is too high, indicating weak spatial-temporal interactions, the generation process will have a high probability of selecting the \MO in order to explore alternative spatial conflict patterns.

\begin{figure}[!t]
    \centering
    \begin{minipage}{0.49\linewidth}
        \centering
        \includegraphics[width=\linewidth]{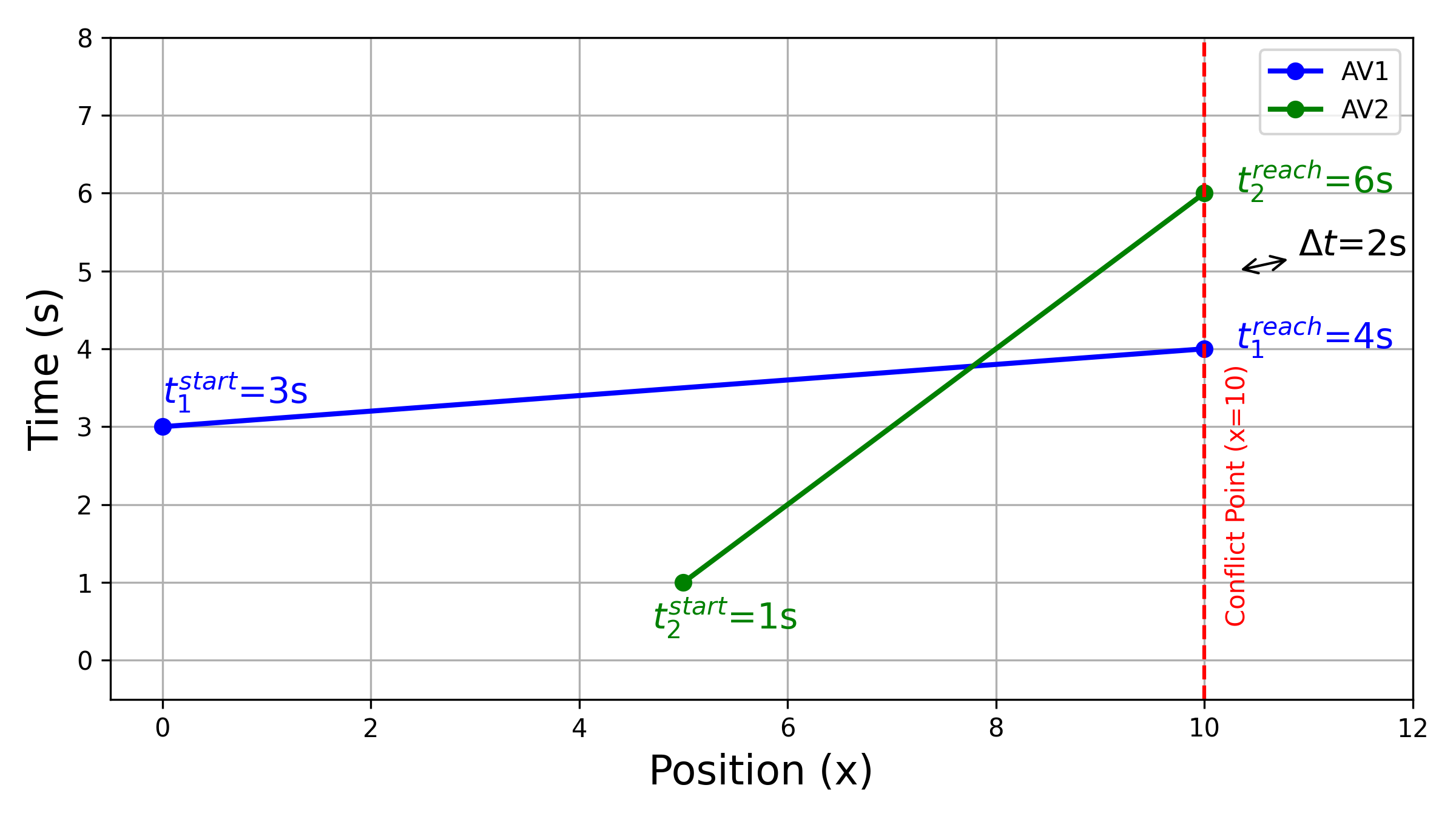}
        \vspace{-20pt}
        \caption{Original scenario.}
        \vspace{-15pt}
        \label{fig:demo-original}
    \end{minipage}
    \hfill
    \begin{minipage}{0.49\linewidth}
        \centering
        \includegraphics[width=\linewidth]{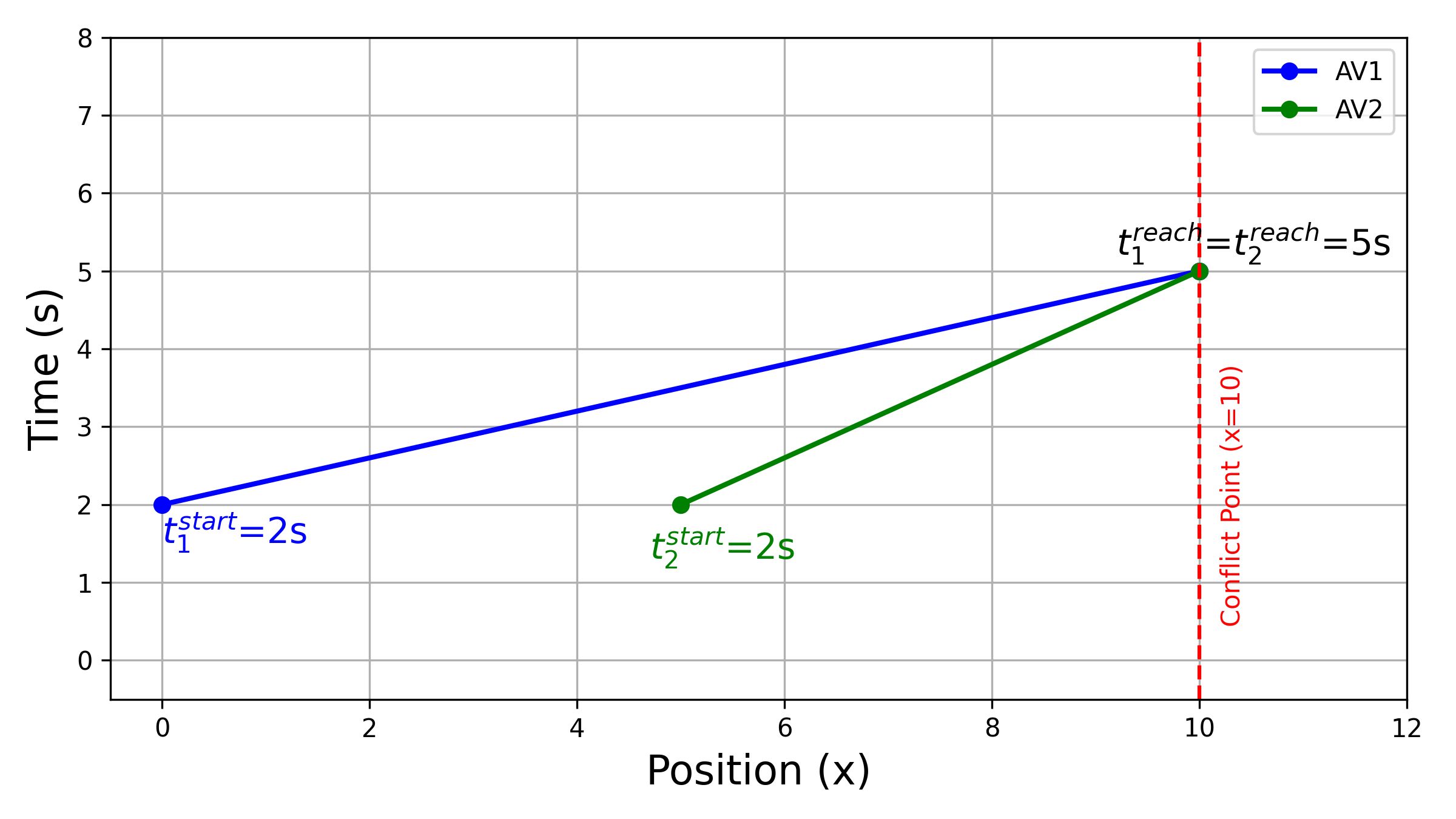}
        \vspace{-20pt}
        \caption{After temporal-aware mutation.}
        \vspace{-15pt}
        \label{fig:demo-change}
    \end{minipage}
\end{figure}

\subsubsection{Temporal-aware Mutation} 
Given a set of AVs \( \mathcal{A} \) and their corresponding conflict point set \( \mathcal{C}_{S} \), the temporal mutation aims to adjust the trigger times of vehicles in \( \mathcal{A} \) to enhance inter-vehicle competitive interactions (Lines~7-16). 
Specifically, a new set \( \mathcal{A}' \) is first initialized as a copy of \( \mathcal{A} \) (Line~10). 
The algorithm then randomly selects a conflict region \( R \in \mathcal{C}_{S} \) and iteratively adjusts the trigger times of the set of AVs that pass through region \( R \), denoted as \( \mathcal{A}_{R} \) (Lines~11--15). 
For each pair \( (\text{AV}_i, \text{AV}_j) \in \mathcal{A}_{R} \), their trigger times are modified to reduce the arrival time gap at \( R \), thereby promoting simultaneous access to the shared region (Lines~12-15). 
% Fig.~\ref{fig:demo-original} and Fig.~\ref{fig:demo-change} illustrate the concept of temporal mutation. As shown in Fig.~\ref{fig:demo-original}, two vehicles, AV1 and AV2, start from different positions and at different times, resulting in their arrival at the conflict point at different timestamps. 
To increase the likelihood of interaction, we temporally adjust the triggering time of both AVs, by the following rule:
\[
t'^{\text{trigger}}_i = \max\left\{0,\ t^{\text{trigger}}_i - \frac{\Delta t'_R}{2} \right\},
\]
where \( \Delta t'_R \) denotes the relative temporal offset between the two AVs at the conflict region. This adjustment aims to synchronize their arrival times so that they reach the conflict point simultaneously in the mutated scenario.
Figs.~\ref{fig:demo-original} and~\ref{fig:demo-change} illustrate the original scenario and the scenario after temporal-aware mutation.
Mutated set \( \mathcal{A}' \) is returned (Line~16).

\subsubsection{Spatial-aware Mutation} 
To prevent the search process from getting stuck in fine-grained optimization within the roads of the current scenarios,
% within a local region of the scenario space, 
we also explore alternative and more competitive spatial conflict configurations among AVs. 
Specifically, given an AV set \( \mathcal{A} \), the spatial mutation begins by initializing a mutated set \( \mathcal{A}' \) as a copy of \( \mathcal{A} \) (Line~18). 
To maintain scenario complexity within reasonable bounds, if the number of AVs exceeds a predefined capacity \( N_{\mathcal{A}} (\geq 3) \), a random subset of size \( n \in [1, N_{\mathcal{A}} - 2] \) is removed (Lines~19-20). 
This ensures that at least two AVs remain to support meaningful multi-AV interactions. 
Next, the spatial mutation performs an offline local search with a maximum of \( N_{\text{local}} \) iterations to identify a new AV configuration that yields a relatively lower estimated spatial-conflict score (Lines~21-28). 
The search begins by randomly initializing a new AV configuration \( \text{AV}_{\text{best}} \), and computing its spatial-conflict score \( \phi^{\text{spatial}}_{\text{best}} \) using the \text{SpatialScore} function (Line~22), as detailed in Section~\ref{sec:st_score}. 
Since the scenario is not executed during this stage, the trajectory of each newly generated AV \( \text{AV}_{\text{tmp}} \) can be estimated using the path planning algorithms applied in the current ADSs, such as the \( \text{A}^* \) algorithm~\cite{thrun2002probabilistic} used in Apollo, Autoware and Roach, based on a start point, destination point, and the road map.
% map's road topology.
In each local search iteration, a new AV configuration \( \text{AV}_{\text{tmp}} \) is randomly sampled (Line~26), and \( \text{AV}_{\text{best}} \) is updated if \( \text{AV}_{\text{tmp}} \) achieves a lower spatial-conflict score (Lines~23-28). 
The best configuration \( \text{AV}_{\text{best}} \) is then selected and incorporated into the scenario to form the mutated AV set \( \mathcal{A}' \) (Line~29). 

In addition, we apply a waypoint mutator~\cite{cheng2023behavexplor, wang2025moditector, av_fuzzer, huai2023sceno} to perturb other non-player character (NPC) participants in the scenario, introducing additional variability and diversity (Line~30). Note that the perturbations are constrained to regions that do not conflict with the AVs, in order to preserve the integrity of the targeted multi-AVs interactions and avoid introducing unintended interference. 
% Additional details are available on our project website~\cite{ourweb}. 

Finally, the mutated AV set \( \mathcal{A}' \) and the participant set \( \mathcal{P}' \) are combined to form the new scenario \( S' = \{ \mathcal{A}', \mathcal{P}' \} \), which is returned for further execution and validation (Lines~5-6).

\section{Evaluation Results}

In this section, we aim to empirically evaluate the capability of \tool on generating ({\dls}s) for ADSs. In particular, we will answer the following research questions:

\noindent \textbf{RQ1:} Can \tool effectively generate {\dls}s for ADSs in multi-vehicle scenarios compared to baselines?

\noindent \textbf{RQ2:} What are the individual contributions of each component in \tool to its overall effectiveness?

\noindent \textbf{RQ3:} How does \tool perform in terms of testing time efficiency?

% \noindent \textbf{RQ4: } Can cooperative mechanism effectively addressing {\dls}s?

% To address these research questions, we conducted experiments using the following settings:

\textbf{Environment.} We performed our experiments using the high-fidelity CARLA simulator~\cite{dosovitskiy2017carla}, evaluating two autonomous driving systems (ADSs): \textit{Roach}~\cite{Renz2022CORL} and \textit{OpenCDA}~\cite{xu2021opencda}. \textit{Roach} is a state-of-the-art end-to-end ADS that directly maps perception outputs to control commands for ego vehicle navigation. In contrast, \textit{OpenCDA} is a modular ADS framework that incorporates cooperative mechanisms among multiple autonomous vehicles. Both systems are fully compatible with the CARLA simulation platform. 
In this paper, our primary objective is to evaluate the decision-making capabilities of ADSs. 
Therefore, following prior studies~\cite{huai2023doppelganger, cheng2023behavexplor, sun2022lawbreaker}, we use ground-truth perception results obtained directly from the simulator to eliminate the influence of perception errors caused by algorithmic inaccuracies or just simulator-induced delays.

\textbf{Driving Scenarios.}  
We evaluate \tool on four representative driving map regions derived from the NHTSA pre-crash typology~\cite{najm2007pre}, which is widely adopted in prior ADS testing research~\cite{wang2025moditector, huai2023doppelganger, av_fuzzer, cheng2023behavexplor}. These regions reflect diverse real-world traffic scenarios that challenge decision-making under multi-vehicle interactions:
\begin{itemize}
    \item \textit{M1: CityRoad Intersection}, a four-way intersection, requiring AVs to handle yielding, turning.
    
    \item \textit{M2: CityRoad T-Junction}, a T-shaped junction where AVs must perform merging or unprotected turns.

    \item \textit{M3: Roundabout}, a circular intersection where AVs must coordinate entry, yielding, and exit.
    
    \item \textit{M4: Highway Merging}, a highway segment with merging lanes, where AVs must safely enter traffic streams.
\end{itemize}
Different from prior testing approaches that rely on fixed route paths for autonomous vehicles (AVs), in this paper, we automatically determine and select both the start point and destination for each AV within the scenario's map region.

\textbf{Baselines.}  
To the best of our knowledge, this is the first work to conduct deadlock testing on ADSs, and thus no existing method serves as a direct baseline for comparison. Therefore, we compare \tool against two representative baselines: (1) \textit{Random} and (2) \textit{DoppelTest}~\cite{huai2023doppelganger}.
Specifically, 
(1) \textit{Random} does not incorporate any feedback or specific mutation strategies. It generates scenarios by randomly deploying AVs and NPC participants with arbitrary configurations.
(2) \textit{DoppelTest} is the only publicly available tool that supports testing with multiple autonomous vehicles. It uses a genetic algorithm guided by multiple safety-critical objectives such as inter-vehicle distance. Although originally designed for the Apollo and SimControl~\cite{apollo}, we reuse its core algorithms to CARLA without modifying the core testing logic.

\textbf{Metrics.} 
To enable a fair comparison with baseline techniques, we integrate our \oracle into each of them to uniformly detect and collect generated \dls{}s. 
In our experiments, we evaluate the effectiveness of \dls{} generation using one primary metric, \#\dls{}, which denotes the number of potential deadlock scenarios identified by \tool. 
In addition, we manually reviewed the simulation recordings corresponding to the detected \dls{}s. 
The number of deadlocks confirmed through human inspection is reported as \#\dls{}-Hum.

% and \#{\dls}-Hum. The metric \#{\dls} denotes the number of potential deadlock scenarios identified by our \oracle.
% To further verify the accuracy of \oracle, all authors independently reviewed the recorded simulation videos corresponding to each detected {\dls}. A scenario is included in \#{\dls}-Hum only if all authors unanimously confirm it as a true deadlock. This dual-metric approach ensures both automated detection coverage and human-verified correctness.

\textbf{Implementation.}  
Similar to previous work~\cite{huai2023doppelganger}, we set the maximum AV capacity \( \mathcal{N}_{\mathcal{A}} \) around 6. 
We empirically set the oracle detection window \( \Delta_{t} \) to 5 seconds based on preliminary observations. The weighting factor $\alpha$ is set to 0.5.
% To reduce non-determinism in simulation-based ADS execution, all experiments are conducted in CARLA’s synchronized mode. 
Each experiment is repeated five times, and we report the average results across runs. For all methods, we allocate a fixed testing budget of two hours per run, which we found adequate for meaningful comparison. 
All experiments are conducted on a platform equipped with an AMD EPYC 9554P CPU, an NVIDIA L40 GPU, and 512GB of RAM. The experiments takes more than 30 GPU days in total.
% Implementation details and source code are available on our website~\cite{ourweb}.

\subsection{RQ1: Effectiveness of \tool}

\begin{table}[!t]
    \centering
    \small
    \caption{Comparison Results with Baselines on OpenCDA.}
    \resizebox{\linewidth}{!}{
    \begin{tabular}{llccccc}
        \toprule
        \multirow{2.5}*{\textbf{ADS}} & \multirow{2.5}*{\textbf{Method}} & \multicolumn{5}{c}{\textit{\#{\dls{}}} (\textit{\#{\dls{}-Hum}})$\uparrow$} \\
        \cmidrule(lr){3-7}
         
         &  & \textit{M1} & \textit{M2} & \textit{M3} & \textit{M4} & \textit{Sum.} \\
        \midrule
        \multirow{3}*{\textit{Roach}} & 
        \textit{Random} & 1.4 (0.6) & 3.2 (2.8) & 0.8 (0.8) & 2.2 (1.2) & \cellcolor{lightgray!30}7.6 (5.4) \\
        
         & \textit{DoppelTest} & 2.8 (2.2) & 2.4 (2.4) & 0.8 (0.2) & 0.2 (0.2) & \cellcolor{lightgray!30}6.2 (5.0) \\
         
         & \tool & \textbf{14.0 (8.0)} & \textbf{24.2 (21.0)} & \textbf{11.6 (7.0)} & \textbf{15.6 (10.4)} & \cellcolor{lightgray!30}\textbf{65.4 (46.4)} \\
         \midrule

         \multirow{3}*{\textit{OpenCDA}} &  \textit{Random} & 7.0 (4.2) & 5.6 (2.4) & 5.6 (3.4) & 19.8 (11.8) & \cellcolor{lightgray!30}38.0 (21.8) \\
         
         & \textit{DoppelTest} & 6.6 (4.2) & 17.2 (7.4) & 6.2 (2.6) & 22.6 (16.2) & \cellcolor{lightgray!30}52.6 (29.2) \\
         
         & \tool & \textbf{28.4 (14.0)} & \textbf{34.0 (10.6)} & \textbf{9.8 (4.0)} & \textbf{50.6 (19.6)} & \cellcolor{lightgray!30}\textbf{122.8 (48.2)} \\
         \bottomrule
    \end{tabular}
    }
    \vspace{-10pt}
    \label{tab:rq1-baseline}
\end{table}

\begin{figure}[!t]
    \centering
    \small
    \includegraphics[width=\linewidth]{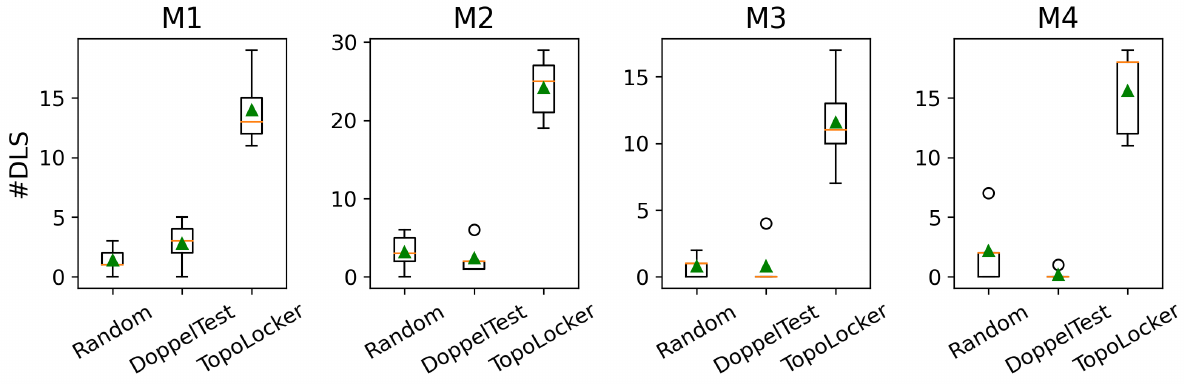}
    \vspace{-20pt}
    \caption{Statistical comparison of \#{\dls{}} on Roach.}
    
    % \caption{Roach Scenario Results}
    \label{fig:rq1-box-roach}
    
    \vspace{0.5em}  % Optional: add vertical spacing between images

     \includegraphics[width=\linewidth]{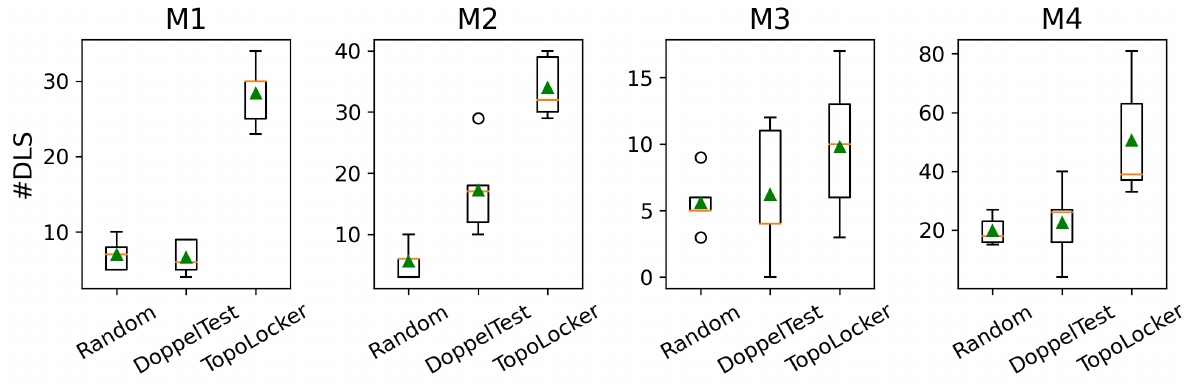}
    \vspace{-20pt}
    \caption{Statistical comparison of \#{\dls{}} on OpenCDA.}
    \vspace{-15pt}
    \label{fig:rq1-box-open}
\end{figure}
\begin{figure}[!t]
    \centering
    \includegraphics[width=0.95\linewidth]{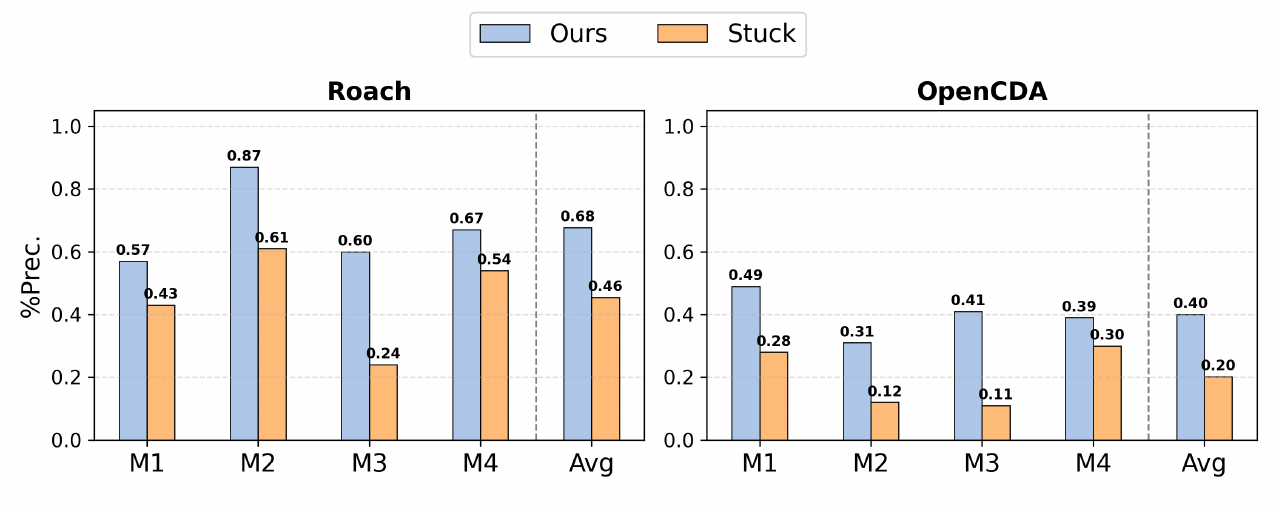}
    \vspace{-10pt}
    \caption{Precision of oracles.}
    \vspace{-10pt}
    \label{fig:rq1-oracle}
\end{figure}
% \begin{table}[!t]
%     \centering
%     \small
%     \caption{Accuracy Comparison with different oracles.}
%     \resizebox{0.75\linewidth}{!}{
%     \begin{tabular}{lccccc}
%         \toprule
%         {\textbf{Oracle}} & \textit{M1} & \textit{M2} & \textit{M3} & \textit{M4} & \textit{Sum.} \\
%         \midrule
%          \textit{Stuck} & \\
%          \textit{Ours} \\ 
%          \bottomrule
%     \end{tabular}
%     }
%     \label{tab:rq1-baseline}
% \end{table}

\subsubsection{Comparative Results} 
Table~\ref{tab:rq1-baseline} compares \#\dls{} and \#\dls{}-Hum across four initial driving scenarios: \textit{M1}, \textit{M2}, \textit{M3}, and \textit{M4}, on two different types of ADSs: \textit{Roach} and \textit{OpenCDA}. 
Fig.~\ref{fig:rq1-box-roach} and Fig.~\ref{fig:rq1-box-open} illustrate the distribution of \#\dls{} across five repeated runs of each method in every scenario, where the orange bar represents the median and the green triangle indicates the average.
Overall, we find that \tool outperforms all baselines in both \#\dls{} and \#\dls{}-Hum across different ADSs and scenarios.

\textit{Roach.} For the end-to-end ADS \textit{Roach}, \tool achieves the best overall performance, significantly outperforming the best baseline (i.e., \textit{Random}) with 65.4 vs. 7.6 detected \#\dls{} cases. 
In detail, \tool surpasses the baseline in all scenarios: \textit{M1} (14.0 vs. 1.4), \textit{M2} (24.2 vs. 3.2), \textit{M3} (11.6 vs. 0.8), and \textit{M4} (15.6 vs. 2.2). 
Among the 65.4 deadlock cases detected by \tool, 46.4 were manually confirmed as valid deadlocks (\#\dls{}-Hum), including 8.0 in \textit{M1}, 21.0 in \textit{M2}, 7.0 in \textit{M3}, and 10.4 in \textit{M4}. 
In contrast, the best-performing baselines achieved only 5.4 (\textit{Random}) and 5.0 (\textit{DoppelTest}) confirmed \#\dls{}-Hum cases.

\textit{OpenCDA.} For the modular ADS \textit{OpenCDA}, which incorporates cooperative communication mechanisms, \tool also demonstrates clear advantages over the baselines. 
It achieves the highest values for both \#\dls{} and \#\dls{}-Hum across all four scenarios. 
Specifically, \tool outperforms the best baseline (i.e., \textit{DoppelTest}) in \textit{M1} (28.4 vs. 6.6), \textit{M2} (34.0 vs. 17.2), \textit{M3} (9.8 vs. 6.2), and \textit{M4} (50.6 vs. 22.6). 
Manual inspection confirms 48.2 valid deadlock cases (\#\dls{}-Hum). 
These results highlight \tool's effectiveness in uncovering challenging multi-AV coordination failures, even in systems equipped with communication and cooperative mechanisms.

\textit{Discussion.} By comparing the overall performance across both ADSs, we observe that fewer deadlocks occur in the \textit{M3} Roundabout scenario. 
This is primarily because roundabouts are relatively rare and less familiar environments for both ADSs, which often struggle to handle them effectively. 
Such unfamiliarity can lead to safety-critical issues or cause the AVs to become stuck without recognizing viable navigational goals.
Additionally, when comparing the overall performance of \textit{Roach} and \textit{OpenCDA}, we find that \textit{Roach}, trained using reinforcement learning, surprisingly demonstrates better cooperative behavior by inducing fewer \dls{}s than \textit{OpenCDA} (65.4 vs. 122.8 \dls{}s detected by \tool). 
A possible explanation is that the cooperative communication and fusion mechanisms in \textit{OpenCDA}, being rule-based, lack generalizability and are highly sensitive to the execution order of decisions among ADSs. 
This makes the system more prone to coordination failures under unfamiliar or adversarial traffic conditions.

% \begin{figure}[!t]
%     \centering
%     \includegraphics[width=0.95\linewidth]{figures/rq1/oracle_prec.pdf}
%     % \vspace{-25pt}
%     \caption{Precision of oracles.}
%     \vspace{-25pt}
%     \label{fig:rq1-oracle}
% \end{figure}
% \begin{figure}[!t]
%     \centering
%     \vspace{-10pt}
%     \includegraphics[width=\linewidth]{figures/case.pdf}
%     \vspace{-55pt}
%     \caption{Case Study of Deadlock Scenarios (DLSs). Solid lines represent execution paths, and dashed lines indicate predicted motion trajectories.}
%     \vspace{-15pt}
%     \label{fig:rq-1-case}
% \end{figure}
\subsubsection{Effectiveness of \oracle}
To evaluate the effectiveness of \oracle, we compare its detection performance against a naive variant that replaces \oracle in \tool with a simple time-based counter for monitoring stuck behavior. 
Fig.~\ref{fig:rq1-oracle} illustrates the precision of both oracles, computed as:
\(
\text{\%Prec.} = \frac{\#\text{DLS-Hum}}{\#\text{DLS}},
\)
which represents the proportion of automatically detected deadlocks that are confirmed through manual inspection.
We observe that our proposed \oracle significantly outperforms the naive time-based oracle, achieving 68\% vs. 46\% on \textit{Roach}, and 40\% vs. 20\% on \textit{OpenCDA}. These results demonstrate that simply detecting stuck behavior is insufficient to capture intention-driven wait-for cycles. 
They also highlight the effectiveness and necessity of our \oracle for accurate deadlock detection in multi-AV scenarios.

\subsubsection{Case Study}
\begin{figure}[!t]
    \centering
    \vspace{-20pt}
    \includegraphics[width=\linewidth]{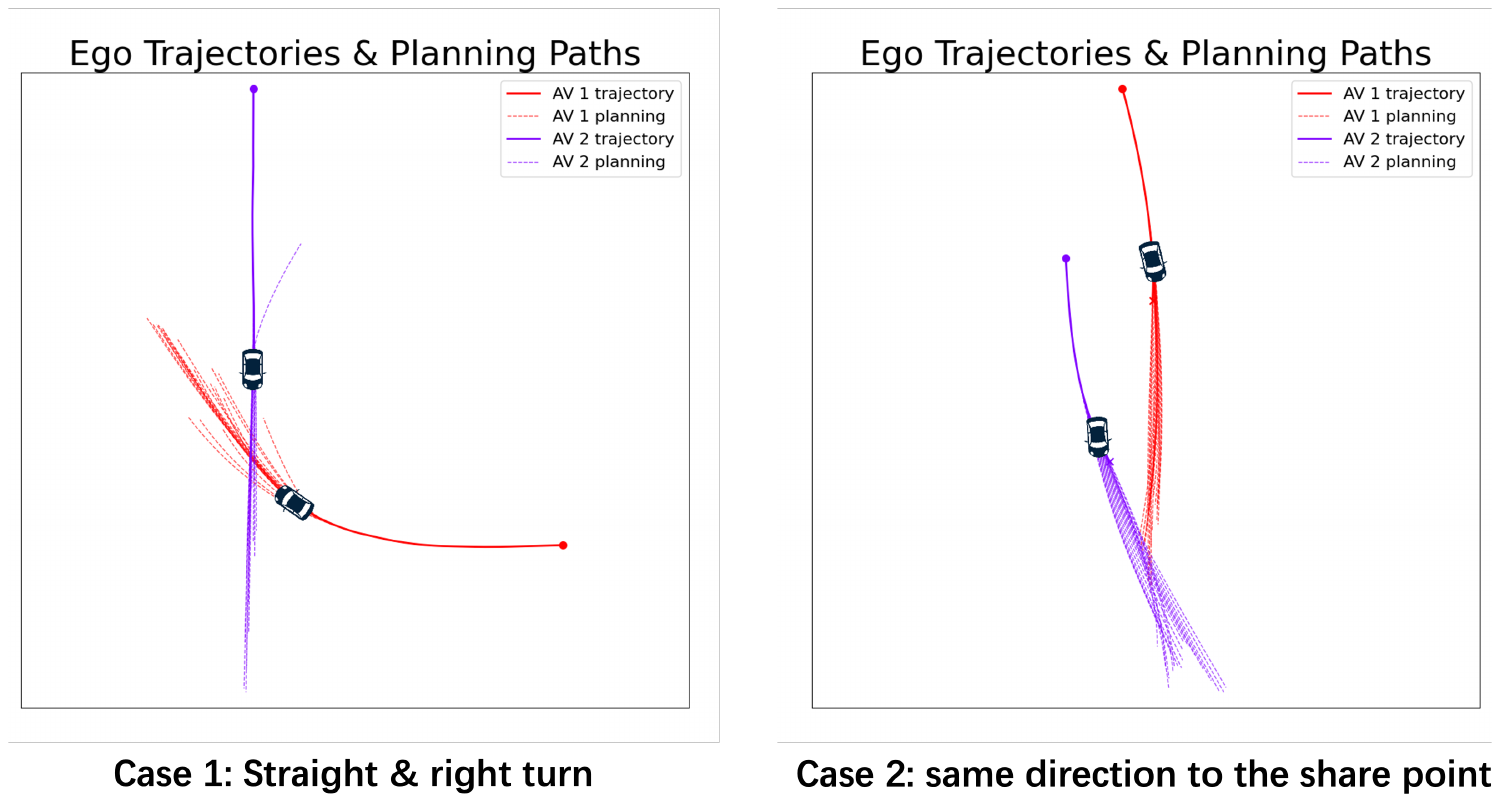}
    \vspace{-55pt}
    \caption{Case Study of Deadlock Scenarios (DLSs). Solid lines represent execution paths, and dashed lines indicate predicted motion trajectories.}
    \vspace{-15pt}
    \label{fig:rq-1-case}
\end{figure}

Fig.~\ref{fig:rq-1-case} shows representative \dls{} examples. 
% Additional examples are available on our website~\cite{ourweb}. 
These cases demonstrate how groups of AVs may fail to coordinate access to shared resources, leading to wait-for states that ultimately result in deadlocks.

\textit{Case 1:} In this scenario, two vehicles, AV1 and AV2, are controlled by Roach to navigate an intersection where AV1 intends to make a right turn while AV2 proceeds straight. As shown in Fig.~\ref{fig:rq-1-case}, both AVs become stuck in a deadlock at the approaching point. During the pre-stop interval, their predicted motion intentions (dash lines) exhibit dense spatial-temporal conflicts, indicating a high likelihood of mutual obstruction.

\textit{Case 2:} This case illustrates a highway merging scenario involving two vehicles, AV1 and AV2, controlled by \textit{OpenCDA}. AV2 slows down to join the platoon led by AV1, while AV1 simultaneously reduces its speed to avoid AV2's motion, which exhibits a high degree of intention overlap with AV1's predicted trajectory. This mutual hesitation results in a deadlock.

% AV1 and AV2 are traveling in the same direction. Their planners independently generate trajectories that pass through a shared point, causing both vehicles to yield to each other and resulting in a deadlock.

% Our results also suggest that cooperative mechanisms may lack robustness in the context of deadlock testing.
\begin{center}
\fcolorbox{black}{gray!10}{\parbox{0.96\linewidth}{\textbf{Answer to RQ1}: \tool significantly outperforms existing methods in identifying \dls{}s under multi-vehicle autonomous driving scenarios. 
}} 
\end{center}
\subsection{RQ2: Usefulness of Components}

\begin{table}[!t]
    \centering
    \small
    \caption{Effectiveness of mutation (top) and feedback (bottom).}
    \vspace{-5pt}
    \resizebox{0.9\linewidth}{!}{
    \begin{tabular}{llccccc}
        \toprule
        \multicolumn{7}{c}{\textbf{(a) Mutation Strategies}} \\
        \midrule
        \textbf{ADS} & \textbf{Method} & \textit{M1} & \textit{M2} & \textit{M3} & \textit{M4} & \textit{Sum.} \\
        \midrule
        \multirow{3}*{\textit{Roach}} & \textit{w/o Temporal} & 8.6 & 14.0 & 5.6 & 9.2 & \cellcolor{lightgray!30}37.4 \\
           & \textit{w/o Spatial} & 11.0 & 15.6 & 9.2 & 10.6 & \cellcolor{lightgray!30}46.4 \\
           & \tool & \textbf{14.0} & \textbf{24.2} & \textbf{11.6} & \textbf{15.6} & \cellcolor{lightgray!30}\textbf{65.4} \\
        \midrule
        \multirow{3}*{\textit{OpenCDA}} & \textit{w/o Temporal} & 23.0 & 27.4 & 9.0 & 40.0 & \cellcolor{lightgray!30}99.4 \\
           & \textit{w/o Spatial} & 32.4 & 30.0 & 6.2 & 46.0 & \cellcolor{lightgray!30}114.6 \\
           & \tool & \textbf{28.4} & \textbf{34.0} & \textbf{9.8} & \textbf{50.6} & \cellcolor{lightgray!30}\textbf{122.8} \\
        \midrule
        \multicolumn{7}{c}{\textbf{(b) Feedback Variants}} \\
        \midrule
        \textbf{ADS} & \textbf{Method} & \textit{M1} & \textit{M2} & \textit{M3} & \textit{M4} & \textit{Sum.} \\
        \midrule
        \multirow{4}*{\textit{Roach}} & \textit{f-Random} & 5.0 & 8.6 & 3.2 & 2.0 & \cellcolor{lightgray!30}18.8 \\
           & \textit{f-Spatial} & 9.0 & 11.4 & 6.0 & 12.2 & \cellcolor{lightgray!30}38.6 \\
           & \textit{f-Temporal} & 11.4 & 12.6 & 4.0 & 7.2 & \cellcolor{lightgray!30}35.2 \\
           & \tool & \textbf{14.0} & \textbf{24.2} & \textbf{11.6} & \textbf{15.6} & \cellcolor{lightgray!30}\textbf{65.4} \\
        \midrule
        \multirow{4}*{\textit{OpenCDA}} & \textit{f-Random} & 16.8 & 18.0 & 5.6 & 23.4 & \cellcolor{lightgray!30}55.2 \\
           & \textit{f-Spatial} & 19.0 & 29.0 & 5.2 & 23.4 & \cellcolor{lightgray!30}76.6 \\
           & \textit{f-Temporal} & 25.4 & 27.0 & 5.6 & 39.0 & \cellcolor{lightgray!30}97.0 \\
           & \tool & \textbf{28.4} & \textbf{34.0} & \textbf{9.8} & \textbf{50.6} & \cellcolor{lightgray!30}\textbf{122.8} \\
        \bottomrule
    \end{tabular}
    }
    \vspace{-15pt}
    \label{tab:rq2-ablation}
\end{table}

We evaluated the key components of \tool's testing process, including the \mutation and \feedback, by configuring a series of variants.

\subsubsection{Mutation}
For the mutation component, we compared \tool{} with two of its ablated variants:  
(1) \textit{w/o Temporal} removes the temporal-aware mutation from \mutation{}, aiming to evaluate the benefits of temporal adjustments in generating deadlock scenarios;  
(2) \textit{w/o Spatial} removes the spatial-aware mutation, allowing scenarios to be generated solely by adjusting the trigger times of ego vehicles. This variant is designed to assess the effectiveness of spatial conflicts in mutation. 

Table~\ref{tab:rq2-ablation} (a) presents the comparison results. We observe that removing either component (\textit{w/o Temporal} or \textit{w/o Spatial}) results in fewer \dls{}s generated compared to the full version of \tool{} across both ADSs. 
Notably, \textit{w/o Temporal} produces the fewest \dls{}, with 37.4 on \textit{Roach} and 99.4 on \textit{OpenCDA}, demonstrating that temporal-aware mutation (i.e., via trigger time adjustment) is particularly effective in uncovering deadlocks. 
This effectiveness stems from the fact that fine-grained timing modifications directly affect the concurrency of AV behaviors at conflict regions, thereby increasing the likelihood of simultaneous contention and deadlock formation.
In addition, \textit{w/o Spatial} also results in performance drops when the spatial-aware mutation is removed: the number of \dls{}s drops from 65.4 to 46.4 on \textit{Roach}, and from 122.8 to 114.6 on \textit{OpenCDA}. This illustrates that spatial mutation contributes by introducing topological conflicts, enabling more competitive configurations to emerge.
Overall, these results demonstrate the complementary contributions of both mutation strategies and highlight the importance of jointly incorporating spatial and temporal mutation to effectively generate deadlock scenarios.

\subsubsection{Feedback}
For the feedback component, we implemented three variants to evaluate its effectiveness:  
(1) \textit{f-Random} replaces the feedback mechanism in \tool{} with random selection, meaning that both seed scenarios and mutation selection are chosen randomly. This variant serves as a baseline to assess the overall value of guided feedback;  
(2) \textit{f-Spatial} relies solely on the \textit{Spatial-Conflict Score} to evaluate its individual contribution;  
(3) \textit{f-Temporal} uses only the \textit{Temporal-Conflict Score} to assess its role in guiding the search.

Table~\ref{tab:rq2-ablation} (b) shows the experimental results in terms of the number of \dls{}s discovered. 
The comparison between \textit{f-Random} and \tool{} (18.8 vs. 65.4 on \textit{Roach}, and 55.2 vs. 122.8 on \textit{OpenCDA}) illustrates the effectiveness of feedback-guided search in \tool. Note that \textit{f-Random} outperforms \textit{Random}, which highlights the contribution of the \mutation{} component even in the absence of any feedback guidance.
Moreover, the results of \textit{f-Spatial} (38.6 on \textit{Roach}, 76.6 on \textit{OpenCDA}) and \textit{f-Temporal} (35.2 on \textit{Roach}, 97.0 on \textit{OpenCDA}) demonstrate that both the spatial and temporal feedback contribute meaningfully to deadlock detection. 
Their combination achieves the best performance.

% we also evaluated the influence of different $\Delta t$ used in non-invasive mutation. Due to the space limit, the detailed experimental results can be found on our website \cite{ourweb}.

\begin{center}
\fcolorbox{black}{gray!10}{\parbox{0.96\linewidth}{\textbf{Answer to RQ2}:
Both the spatial and temporal designs in the mutation and feedback components are essential for enabling \tool{} to effectively generate deadlock scenarios.
 }}
\end{center}
\subsection{RQ3: Testing Performance}

\begin{table}[!t]
    \centering
    \small
    \caption{Result of time performance (second).}
    \vspace{-5pt}
    \resizebox{\linewidth}{!}{
    \begin{tabular}{lccccc}
        \toprule
        \textbf{Method} &  \textbf{Mutation} & \textbf{Oracle} & \textbf{Feedback} & \textbf{Simulation} & \textbf{Total} \\
        \midrule
         \textit{Random} & 3.99 & N/A & N/A  & 52.87 & 56.86 \\
        
         \textit{DoppelTest} & 4.18 & 0.08 & 0.37 & 45.27 & 49.90 \\
         
         \tool &7.62 & 0.72 & 1.18 & 61.89 & 71.41 \\
         \bottomrule
    \end{tabular}
    }
    \vspace{-15pt}
    \label{tab:rq3-time}
\end{table}

We further assess the time performance of the main components in \tool, including the overhead introduced by mutation, oracle checking, feedback computation, and simulation execution. 
Specifically, we analyze the average time required to process a single scenario. 
The results are summarized in Table~\ref{tab:rq3-time}, where the columns \textit{Mutation}, \textit{Oracle}, \textit{Feedback}, and \textit{Simulation} report the average time (in seconds) spent on each corresponding stage: applying mutations, performing oracle checks, computing feedback scores, and running the scenario in the simulation platform. 
% Values marked as \texttt{0.01*} indicate negligible overhead (i.e., less than 0.01 seconds). 
For the \textit{Random} variant, the oracle and feedback stages are not applicable and are marked as `N/A'.

Overall, the results show that simulation dominates the total processing time in all methods. 
For example, on average, \tool takes 71.41 seconds to process a single scenario, of which 61.89 seconds (86.7\%) are spent on simulation. 
This higher simulation time, compared to other approaches, is primarily due to the need to observe a prolonged stationary period for effective deadlock detection. 
In terms of testing components, \tool incurs slightly more time in mutation, oracle evaluation, and feedback processing, reflecting the additional computation required to guide the generation of deadlock-prone scenarios. 
The most time-consuming operations are the spatial conflict pre-checking in the spatial-aware mutation and the conflict region identification during feedback evaluation.
Nevertheless, simulation time remains the dominant component of overall testing, and the additional overhead introduced by \tool has only a minor impact on total testing time and overall testing performance.

\begin{center}
\fcolorbox{black}{gray!10}{\parbox{0.96\linewidth}{\textbf{Answer to RQ3}:\tool slightly increases the proportion of non-simulation computation time; however, the majority of the testing time is still dominated by simulation.}}
\end{center}

\subsection{Threats to Validity}
\tool{} faces several potential threats to validity. First, the choice of the autonomous driving systems (ADSs) under evaluation may influence the results. To mitigate this concern, we conducted experiments on two distinct types of ADSs that are widely used, well-evaluated, and stably compatible with the CARLA simulator. Since our method is ADS-agnostic by design, it can be readily extended to other ADSs, which we plan to explore in future work. 
Second, our evaluation does not consider the effects of perception errors. As this paper focuses specifically on assessing the decision-making modules of ADSs, we follow prior work~\cite{cheng2023behavexplor, huai2023doppelganger, sun2022lawbreaker} in excluding perception-related noise, which could otherwise obscure the clarity of our analysis.
Third, the accuracy of wait-for graph construction and the configuration of stationary period thresholds may pose potential threats to validity. In particular, inaccurate motion prediction could lead to incorrect intention inference, and premature classification of deadlocks may occur if ego vehicles eventually resume movement. To mitigate these risks, we manually reviewed each case and excluded those exhibiting significant prediction errors or false deadlock indications, ensuring the reliability of our analysis.
Lastly, the diversity of driving scenarios and the inherent non-determinism in ADS execution may also pose threats to the validity of our results. To mitigate these factors, we select a wide range of different driving scenarios and repeat each experiment multiple times to account for stochastic variations.

\section{Related Works}
\textbf{ADS Testing.} 
As testing is critical for ensuring the safety and reliability of autonomous driving systems (ADS), the development of efficient and effective testing methodologies has garnered sustained attention from the research community. Broadly, existing approaches can be categorized into two major groups: data-driven testing~\cite{zhang2023building,deng2022scenario,gambi2019generating,najm2013depiction,nitsche2017pre,roesener2016scenario,paardekooper2019automatic,lu2024diavio} and search-based testing~\cite{cheng2023behavexplor,han2021preliminary,av_fuzzer,icse_samota,tse_adfuzz,tang2021systematic,zhou2023specification,tang2021route,tang2021collision,huai2023doppelganger,wang2025moditector,tang2025moral}. 
\textit{Data-driven testing} leverages real-world data sources to construct critical test scenarios. These sources include traffic logs~\cite{deng2022scenario,paardekooper2019automatic,roesener2016scenario,tang2023evoscenario} and accident reports~\cite{gambi2019generating,najm2013depiction,nitsche2017pre,zhang2023building,guo2024sovar,tang2024legend}, enabling the reproduction or simulation of historically risky driving behaviors.
\textit{Search-based testing}, on the other hand, explores the scenario space systematically to identify safety-critical cases. Techniques in this category include:
guided fuzzing~\cite{cheng2023behavexplor,av_fuzzer,MDPFuzz_2022_issta},
evolutionary algorithms~\cite{gambi2019automatically,han2021preliminary,tang2021collision,tang2021route,tang2021systematic,zhou2023specification,tian2022mosat},
metamorphic testing~\cite{han2020metamorphic},
surrogate model-guided search~\cite{icse_samota,tse_adfuzz},
reinforcement learning~\cite{haq2023many,feng2023dense,lu2022learning}, 
and reachability analysis~\cite{hildebrandt2023physcov,althoff2018automatic}.
These methods have demonstrated their effectiveness in revealing vulnerabilities in ADS decision-making. However, most focus on either individual vehicle behavior or single-agent criticality, with limited attention paid to multi-agent interactions such as deadlock, a gap that our work seeks to address.

\textbf{Cooperative Autonomous Driving.}
Cooperative autonomous driving is a paradigm in which two or more connected automated vehicles (CAVs) communicate and coordinate their actions to achieve their respective driving objectives~\cite{automation2020taxonomy}. This paradigm aims to enhance the efficiency, safety, and reliability of ADSs in complex traffic environments.
Existing approaches to cooperative driving can be broadly classified into two categories. 
The first involves \emph{decentralized cooperation}, where individual vehicles exchange information such as sensor data~\cite{thandavarayan2020cooperative}, perception results~\cite{chiu2025v2vllmvehicletovehiclecooperativeautonomous,wang2020v2vnet}, or planning outputs~\cite{yan2024multi,yan2023unified}. 
This model enables vehicles to adapt their behavior in real time based on local observations and peer communications. 
However, it also introduces challenges related to communication reliability and the complexity of multi-source data fusion~\cite{gao2024survey,ngo2023cooperative}.
The second category is \emph{centralized coordination}, where a central controller (or master) plans and distributes decisions to all participating vehicles~\cite{gholamhosseinian2022comprehensive,chen2021hierarchical}. 
While centralized systems offer more stable and globally optimized coordination, they typically suffer from limited adaptability and increased latency in dynamic environments.
Although cooperative autonomous driving has gained growing attention, research on multi-vehicle collaboration remains relatively limited compared to the extensive work on individual ADSs~\cite{ji2024Toward}. In particular, the evaluation of failure modes such as coordination conflicts and deadlock scenarios has not been sufficiently explored. Our work also addresses this gap by systematically testing cooperative ADS behaviors under conditions that may lead to inter-vehicle deadlocks.

\section{Conclusion}
In this paper, we present the first study to evaluate the cooperative capabilities of ADSs, with a particular focus on their ability to handle deadlocks in multi-vehicle traffic scenarios. 
To address this challenge, we develop an effective testing tool, \tool, which integrates three key components: \oracle, to accurately identify wait-for deadlock situations among multiple autonomous vehicles; \feedback and \mutation, which work collaboratively to guide the search process to discover more deadlock scenarios. 
Experimental results demonstrate the effectiveness and efficiency of \tool, as well as the contribution of each individual component.

% \input{sections/data_available}

\bibliographystyle{ACM-Reference-Format}
\bibliography{reference}

\end{document}